\documentclass[11pt]{article}
\usepackage[T1]{fontenc}
\usepackage{amssymb,amsmath,amsthm,amsfonts,dsfont,bm}
\usepackage{multirow}
\usepackage{graphicx}
\usepackage[hang,small,bf]{caption}
\usepackage{lscape}
\usepackage{color}
\usepackage{hyperref,url}
\usepackage[round,sort,comma]{natbib}
\newcounter{bibcount}
\usepackage{enumitem}
\usepackage[title]{appendix}
\usepackage{float}
\usepackage{mathtools}
\DeclarePairedDelimiter\ceil{\lceil}{\rceil}
\DeclarePairedDelimiter\floor{\lfloor}{\rfloor}
\usepackage[english]{babel}
\usepackage[autostyle, english = american]{csquotes}
\MakeOuterQuote{"}

\usepackage{silence}
\newcommand{\E}{\mathbb{E}}
\newcommand{\Var}{\mathbb{V}ar}
\newcommand{\Cov}{\mathbb{C}ov}

\definecolor{darkblue}{rgb}{0,0,0.55}

\patchcmd{\appendices}{\quad}{: }{}{} 

\newtheorem{thm}{Theorem}
\newtheorem{lemma}{Lemma} 

\numberwithin{thm}{section}
\numberwithin{defn}{section}

\numberwithin{equation}{section}
\numberwithin{figure}{section}
\numberwithin{table}{section}
\newtheorem{note}{Note}[section]

\makeatletter
\patchcmd{\@lbibitem}%
{\item[}%
{\item[\hfill\stepcounter{bibcount}{[\thebibcount]}}%
{}%
{}
\patchcmd{\@lbibitem}%
{\hfil \NAT@anchor {#2}{\NAT@num }]}%
{\hyper@natanchorstart{#2\@extra@b@citeb}\hyper@natanchorend]}
{}%
{}
\renewcommand\NAT@bibsetup[1]{%
	\setlength{\leftmargin}{0.40in}
	\setlength{\itemindent}{-\parindent}%
	\setlength{\itemsep}{\bibsep}%
	\setlength{\parsep}{\z@}}
\makeatother

\hypersetup{,colorlinks=true,allcolors=darkblue}

\allowdisplaybreaks

\begin{document}   

\baselineskip 5mm

\thispagestyle{empty}

\begin{center}
{\LARGE Credibility Theory Based on  Winsorizing
}

\vspace{12mm}

{\large\sc
Qian Zhao\footnote[1]
{
{\sc Corresponding Author:}
Qian Zhao, Ph.D., ASA, is
an Associate Professor in the Department of Mathematics, 
Robert Morris University, Moon Township, PA 15108, USA. 
~~ {\em E-mail\/}: ~{\tt zhao@rmu.edu}}}  

\vspace{1mm}

{\large\em Robert Morris University}

\vspace{8mm}

{\large\sc
Chudamani Poudyal\footnote[2]{ 
Chudamani Poudyal, Ph.D., ASA, 
is an Assistant Professor 
in the Department of Mathematical Sciences, 
University of Wisconsin-Milwaukee, 
Milwaukee, WI 53211, USA. 
~~ {\em E-mail\/}: ~{\tt cpoudyal@uwm.edu}}}

\vspace{1mm}

{\large\em University of Wisconsin-Milwaukee}

\end{center}

\vspace{2mm}

\begin{quote}
{\bf\em Abstract\/}.
The classical B\"{u}hlmann credibility model has been 
widely applied to premium estimation for group insurance 
contracts and other insurance types. 
In this paper, we develop a robust B\"{u}hlmann credibility
model using the winsorized version of loss data, 
also known as the winsorized mean 
(a robust alternative to the traditional individual mean).
This approach assumes that the observed sample data come 
from a contaminated underlying model with 
a small percentage of contaminated sample data. 
This framework provides explicit formulas for the 
structural parameters in credibility estimation 
for scale-shape distribution families, 
location-scale distribution families,
and their variants, 
commonly used in insurance risk modeling.
Using the theory of \(L\)-estimators 
(different from the influence function approach), 
we derive the asymptotic properties of the proposed
method and validate them through a comprehensive simulation study,
comparing their performance to credibility based on the trimmed mean. 
By varying the winsorizing/trimming thresholds in several parametric models,
we find that all structural parameters derived from the 
winsorized approach are less volatile than those from the trimmed approach. 
Using the winsorized mean as a robust risk measure can 
reduce the influence of parametric loss assumptions 
on credibility estimation. 
Additionally, we discuss non-parametric estimations in credibility.
Finally, a numerical illustration from the
Wisconsin Local Government Property Insurance Fund 
indicates that the proposed robust credibility approach
mitigates the impact of model mis-specification
and captures the risk behavior of loss data
from a broader perspective.

\vspace{20pt}

\end{quote}
\newpage

\section{Introduction}
\label{sec:Introduction}

The B\"{u}hlmann model is a classical linear credibility model that is successively applied at premium estimation and other insurance specifications. The central idea is: for a particular policyholder, we have observed $n$ exposure units of past claims ${\bf{X}}=(X_{1}, X_{2}, \hdots, X_{n})$. Assume these losses follow the common risk parameter $\Theta$ (also a random varible) and $X_{1}|\Theta, X_{2}|\Theta, \hdots, X_{n}|\Theta$ are independent
and identically distributed (i.i.d.) on condition $\Theta= \theta$, then the ideal manual rate (we call it hypothetical mean), for an insured with $\theta$ is $\mu(\theta)= \E[X|\theta]$.
\cite{buhlmann_1967} determined
the linear credibility premium by minimizing the expected squared loss:
\begin{align}
\label{eqn:BhulmannCredibility1}
\min_{\alpha, \beta} 
\E
\left[
\left(
\mu(\theta)-\alpha- \beta \Bar{X}
\right)^{2}
\right]
\quad 
\mbox{where}
\quad 
\Bar{X}
=
n^{-1} 
\sum_{i=1}^{n} 
X_{i}.
\end{align}
The resulting credibility premium is 
$$P_{c}= z\bar{X}+(1-z)\mu,\qquad z=\frac{n}{n+v/a},$$
where $\mu=\E[\mu(\theta)]$ 
is the collective premium for the portfolio of all insureds, 
$n$ is the size of data, 
$v=\E[\Var(X|\theta)]$ and $a=\Var(\mu(\theta)) $. \\

In the last several decades, various adaptations and extensions have been made in the credibility literature following the B\"{u}hlmann’s approach.
One of the extensions yet to be investigated 
is the robust method in the area of credibility. Using conventional mean $\Bar{X}$ in (\ref{eqn:BhulmannCredibility1}), excess claims could have significant distorting effects and lead to an unsatisfactory behavior of standard linear credibility in estimators. To overcome this, a number of authors have studied the combination of credibility and robust statistics in order to obtain better risk management. 
\cite{kunsch_1992} proposed the linear credibility estimator with the claims replaced by a robust $M-$ estimator. 
\cite{gisler_reinhard_1993} investigated the application of robust credibility in the general $\textrm{B\"{u}hlmann-Straub}$ model. \cite{Buhlmann2005} discussed semilinear credibility with truncation and weights. Their main idea is to consider the correlation between the original data and their transformed data in credibility estimation. However, this research focuses on pure robust credibility based on the transformed data itself and derives an explicit formula for asymptotic properties.
\cite{db07} further refined robust procedures
by reducing the degree of heuristic arguments 
for the credibility premium when claims
are approximately gamma-distributed. Besides, \cite{Garrido2000} extended the study by focusing on the bias-treatment alternatives with the robust portfolio-unbiased procedure and pure robust credibility. The asymptotic optimality of pure robust credibility is also proved in \cite{PITSELIS20132}.  These scholar works have the same main idea -- that is to robustify the claims experience by using an alternative risk measure instead of the traditional individual mean.\\

The intersection of robust credibility and linear models was also investigated in many literature.  \cite{PITSELIS2008}  applied robust statistics to the regression credibility estimation by using the influence function approach of $M$-estimators. \cite{Dornheim2011a,Dornheim2011b} and \cite{Brazauskas2013} introduced a truncation likelihood-based approach for robust-efficient fitting of mixed linear models.
These methods yield more accurate premiums when extreme outcomes are present in the data and the procedures are flexible and effective risk-pricing tools in the fields of property and casualty insurance, health care, and real
estate. Such robust estimators can also be applied to the credibility theory for generalized mixed linear models
\citep[see, e.g.,][]{nelder_verrall_1997, Garrido2006, Lai_Sun2012, Er2016}. Additionally, when the sample mean or the
individual-level claim data are not available in \eqref{eqn:BhulmannCredibility1}, \cite{MR4397761} developed a general optimally weighted approach for mixed credibility to overcome the scenario.
\\

Regarding the performance of various robust estimators applying in credibility and ratemaking,
\cite{Gisler1980} discussed the optimum trimming of data in the credibility framework.
\cite{PITSELIS2013} compared multiple robust estimators in regression credibility.
With a trimmed mean as the risk control,
\cite{Kim2013} investigated the general asymptotic properties of the structural parameters in credibility theory. 
This pure robust credibility in the 
B{\"u}hlmann-Straub
model can be easily adapted to an insurance contract specifications and handle extreme losses. 
However, this approach discards all information contained in outlying data points even though they might be legitimate
observations from the actual assumed loss model.
To address this shortcoming, our contribution 
in this scholarly work is to propose a credibility 
theory based on winsorizing the original sample 
data to mitigate the impact of outliers or model 
mis-specification in credibility premium estimation.
Specifically, ground-up losses are winsorized 
from above and below with different proportions,
for example, with probabilities of $100p\%$ and $100q\%$, 
respectively, 
to examine how winsorized-robust credibility 
structural parameters compare to corresponding 
non-robust quantities.
Asymptotic properties of this robust credibility 
estimators are established 
following the general $L$-statistics approach 
developed by \cite{Chernoff1967} 
rather than influence function approach 
used for trimmed version by \cite{Kim2013}.
Simulation studies are used to illustrate
the advantage and disadvantage of this 
approach compared with the method based 
on trimming the data. \\

The remainder of this paper is organized as follows.
In Section \ref{sec:RobustRisk},
we present the idea of winsorized mean as a risk measure. 
Section \ref{sec:CredRobustData}
develops the credibility theory and the 
corresponding asymptotic properties based
on winsorizing the data. 
Parametric model examples o
f credibility premiums for two types of family 
distributions -- scale-shape 
family and location-scale family are provided 
in Section \ref{sec:ParametricExamples}. 
The effect of model choice and parameter estimation 
method on credibility premium is illustrated through 
a comprehensive simulation study 
in Section \ref{sec:Sim}.
In Section \ref{sec:NonPar}, 
we demonstrate how the proposed approach can be 
used in the non-parameter setting. 
We conclude the paper with a brief summary
of the main findings in Section \ref{sec:Conclusion}.

\section{Robust Mean as the Risk Measure}
\label{sec:RobustRisk}

In Section 2.1, we outline the idea of trimming and winsorizing data and show their robustness to outliers that often occur in the risk models. We defined the random variable for trimmed mean and winsorized mean, respectively. 
The properties of the trimmed mean as the risk measure have been examined in \cite{Kim2013} and the counterparts for winsorized mean are discussed in Section 2.2. The asymptotic properties of these two risk measures are detailed in \cite{Brazauskas2009} and \cite{zhao_brazauskas_ghorai_2018}  and further described in Section 3.

\subsection{Idea of Trimming and Winsorizing}

In actuarial science, loss distributions often can
be heavily influenced by extreme values (outliers). To overcome this issue, statisticians consider the strategy of trimming and winsorizing, which are typically transformations of data by
limiting extreme values in the data set, to reduce the effect of possibly spurious outliers (see \cite{Huber1980}).
The trimming is to exclude all outliers, whereas the winsorizing is to set all outliers to a specified percentile of the data.
The estimators based on the trimmed and winsorized data are usually more robust to outliers than their standard counterparts and commonly used in data analysis for life and non-life insurance products.\\

Now, we see how the trimming or winsorizing strategy work in the credibility model.
Consider a loss random variables $X$ with a cumulative distribution function $F(x|\Theta)$. The parameter $\Theta=\theta$ is a (prior) random variable with density $\pi(\theta)$. We denote the corresponding quantile function for $X|\theta$ as $F^{-1}(w;\theta)$ (to make things easy, in this paper, we use $F^{-1}(w)$ instead). This leads to 
a transformed random variable $X_{T}|\theta$, with $100p\%$ left-trimmed and $100q\%$ right-trimmed data, denoting as 
\begin{equation}
 X_{T}|\theta
 =
\begin{cases}
F^{-1}(w)=x; & \text{if}\quad  p\leq w\leq 1-q ,\\
0; & \text{otherwise}, \\
\end{cases}   
\label{eqn: T}
\end{equation}
and another winsorized random variable $X_{W}|\theta$, with $100p\%$ left-winsorized and $100q\%$ right-winsorized data, denoting as 
\begin{equation}
 X_{W}|\theta
 =
\begin{cases}
F^{-1}(p)=x_{p}; & \text{if} \quad w<p, \\
F^{-1}(w)=x; & \text{if}\quad  p\leq w\leq 1-q, \\
F^{-1}(1-q)=x_{1-q}; & \text{if}\quad w> 1-q, \\
\end{cases} 
\label{eqn: W}
\end{equation}
where $0\leq p<1-q\leq 1$.\\

Then the trimmed and winsorized $k$ moments are derived as 
\begin{subequations}
\begin{align}
\E[X_{T}^{k}|\theta]
&= \frac{1}{1-p-q}
\int_{p}^{1-q}  [F^{-1}(w)]^{k} \,du,\label{eqn:moment_T}   \\
\E[X_{W}^{k}|\theta]
& = p^{k} \left[F^{-1}(p)\right]^{k} +
\int_{p}^{1-q}  [F^{-1}(w)]^{k} \,du +
q^{k} \left[F^{-1}(1-q)\right]^{k},
\label{eqn:moment_W}
\end{align} 
\end{subequations}
  
\noindent
where the proportions $p$ and $q$ can be controlled by the researcher. If we set  $X_{R}|\theta$ as a general robust random variable for (\ref{eqn: T}) and (\ref{eqn: W}), then $\mu_{R}(\theta)=\E[X_{R}|\theta]$ reflects theoritical trimmed mean or winsorized mean with a specific robust type $R\in \{T,W\}$. The choice of $p$ and $q$ has a significant effect on this risk control and further credibility premium in Section 4 and thereafter.

\subsection{Winsorized Mean as a Risk Measure}
In loss models, both the trimmed mean and winsorized mean can be used to quantify the risk exposures.
\cite{Kim2013} have already examined the coherent properties corresponding to the trimmed mean. 
Now we use a similar procedure to discuss the counterparts but for winsorized mean 
--
the alternative and more robust risk measure. \\

According to \cite{MR1892189} and \cite{MR1929369},
a coherent risk measure satisfies
four desirable properties of 
subadditivity,
monotonicity,
positive homogeneity, and 
translation invariance.
Next, we check if the proposed risk 
control measure possesses all 
the properties/axioms in the following theorem.

\medskip 

\begin{thm}
Let $\rho(X)$ denotes the winsorized mean 
as defined in Eq. \eqref{eqn:moment_W} and for $k=1$. 
For $0 < p < 1-q < 1$, $\rho(X)$
satisfies all the coherent risk measure 
axioms except for the subadditivity.  

\begin{proof}
For notational simplicity, 
we suppress the underlying parameter $\theta$. 
Let $X$ and $Y$ be two continuous loss random variables. 
Now, we investigate each of the axioms of a 
coherent risk measure one at a time. 

\begin{enumerate}[label=(\roman*)]
\item
{\sc Subadditivity:} 
This axiom does not hold. 
For that, following \cite{Kim2013},
we give a counter example. 
Let $X$ and $Y$ be both standard normal
but independent random variables. 
Following \cite{MR1956509, MR1932751},
let $\left( X^{c}, Y^{c} \right)$
be the comonotonic part of 
$(X,Y)$.
Define $Z = X+Y$ and $Z^{c} = X^{c}+Y^{c}$.
Then, clearly
$Z = X+Y \sim N(0, (\sqrt{2})^{2})$
and
$Z^{c} = X^{c}+Y^{c} \sim N(0, 2^{2})$. 
Since the distribution functions
of $Z$ and $Z^{c}$ only cross once 
at $(0,0.5)$ then for 
$0 < u < 0.5$, we have 
\begin{align}
\label{eqn:SubAdd1}
F_{X+Y}^{-1}(u)
& =
F_{Z}^{-1}(u) 
> 
F_{Z^{c}}^{-1}(u) 
=
F_{X}^{-1}(u)
+
F_{Y}^{-1}(u).
\end{align}
Thus, for any winsorizing proportions $p$ and $q$ 
satisfying 
$
0 < p < 1-q< 0.5,$
we get
\begin{align}
\rho(X+Y) 
& = 
\rho(Z)
\nonumber \\ 
& = 
p  F_{Z}^{-1}(p) 
+
\int_{p}^{1-q} 
F_{Z}^{-1}(u) \, du 
+
q  F_{Z}^{-1}(1-q) 
\nonumber \\ 
& >
p  F_{Z^{c}}^{-1}(p) 
+
\int_{p}^{1-q} 
F_{Z^{c}}^{-1}(u) \, du 
+
q  F_{Z^{c}}^{-1}(1-q) 
\nonumber \\ 
& = 
p
\left[ 
F_{X}^{-1}(p)
+
F_{Y}^{-1}(p)
\right]
+ 
\int_{p}^{1-q} 
\left[
F_{X}^{-1}(u)
+
F_{Y}^{-1}(u)
\right] \, du 
+ 
q
\left[ 
F_{X}^{-1}(1-q)
+
F_{Y}^{-1}(1-q)
\right] 
\nonumber \\ 
& = 
\rho(X) 
+ 
\rho(Y) 
\nonumber \\
\mbox{i.e.,} \quad 
\rho(X+Y) 
& >
\rho(X) 
+ 
\rho(Y).
\label{eqn:SubAdd2}
\end{align}

\item 
{\sc Monotonicity:}
If $\Pr(X\leq Y)=1$ always holds, 
then so does $F_{X}(t)\geq F_{Y}(t)$ for any $t$. 
Hence, $F_{X}^{-1}(u)\leq F_{Y}^{-1}(u)$ for any $0<u<1$. 
Then
\[
p  F_{X}^{-1}(p) +
\int_{p}^{1-q}  F_{X}^{-1}(u) \,du +
q  F_{X}^{-1}(1-q)\leq p  F_{Y}^{-1}(p) +
\int_{p}^{1-q}  F_{Y}^{-1}(u) \,du +
q  F_{Y}^{-1}(1-q).
\]
Thus, $\rho(X)\leq \rho(Y)$.

\item
{\sc Positive Homogeneity (scale equivalent):} 
For any positive constant $c$,
by equations (2.1) and (2.2), 
we have 
\[ 
\rho(c X)
= 
p\,cx_{p} +
\int_{p}^{1-q}cx \,du +q\,  cx_{1-q}=c\Big(p x_{p} 
+
\int_{p}^{1-q}x \,du +q  x_{1-q}\Big)=c\rho(X).
\]

\item 
{\sc Translation Invariance:}
For any constant $c$, 
we have 
\begin{align*}
\rho(X+c)
& =
p (x_{p}+c) 
+
\int_{p}^{1-q}(x+c) \, du 
+ 
q  (x_{1-q}+c) \\
& =
\left( 
p x_{p} 
+
\int_{p}^{1-q}x \, du 
+
q  x_{1-q}
\right)
+ 
c
\left(
p +
\int_{p}^{1-q}1 \,du + q 
\right) \\ 
& =
\rho(X) + c.
\end{align*}
\end{enumerate}
Thus, the winsorized mean satisfies all the 
coherent axioms except for the subadditivity.
\qedhere 
\end{proof}
\end{thm}

\section{Credibility Based on Robust Data}
\label{sec:CredRobustData}

Let $X_{1}|\theta, \hdots, X_{n}|\theta$ be i.i.d. random variables with 
a common parametric distribution $F(x|\theta)$. 
Denote the order statistics of 
$X_1, \ldots, X_n$ by $X_{(1)} \le \hdots \le X_{(n)}$. 
To make our notation consistent with that of the 
$\textrm{B\"{u}hlmann}$ credibility theory, 
we define the following structural parameters
of robust random variables with 
specific proportions $p$ and $q$, and $R \in \{T,W\}$.
\begin{itemize}
\item 
Collective premium 
$\mu_{R}=\E[\E[X_{R}|\theta]]=\E[\mu_{R}(\theta)]$ 
with $\mu_{R}(\theta)$ is defined in (\ref{eqn:moment_T}) or (\ref{eqn:moment_W}).

\item Expectation of process variance
$v_{R}=\E[v_{R}(\theta)]$ with $v_{R}(\theta)$
is defined in (\ref{eqn:var_T}) or (\ref{eqn:var_W}).

\item Variance of hypothetical means $a_{R}=\Var(\mu_{R}(\theta))$.
\end{itemize}

\noindent
The robust B{\"{u}}hlmann credibility models presented in this paper
are essentially linear approximations of the corresponding 
Bayesian credibility models.\label{page:ExpGamma1}
Based upon (\ref{eqn:moment_T}) and (\ref{eqn:moment_W}), the empirical version of the robust mean (first moment) for a vector ${\bf{X}}
=
\left(X_{1},\hdots,X_{n}\right)$ is 
\begin{equation}
 \widehat{R}({\bf{X}}) =
\begin{cases}
\widehat{T}({\bf{X}}) = \dfrac{1}{n-[n p]-[n q]} 
\sum_{i=[n p]+1}^{n-[n q]}  X_{(i)}, \\[1.25ex]
       \widehat{W}({\bf{X}}) = \dfrac{1}{n} 
\left[ 
 [n p] \, X_{( [n p]+1)}  +
\sum_{i=[n p]+1}^{n-[n q]}  X_{(i)} +
[n q]\,  X_{(n - [n q])})  \right],
\end{cases}
\label{eqn: robust}
\end{equation}
where $[\cdot]$ denotes the greatest integer part. \\

By the structure of (\ref{eqn:BhulmannCredibility1}), a classic approach to determine the credibility premium with trimmed data or winsorized data is minimizing the expected square loss 
\begin{equation}
\underset{\alpha, \beta}{\text{min}} 
\quad  
\E \bigg\{ \big[\mu_{R}(\theta)-(\alpha\,+\beta \, \widehat{R}({\bf{X}})\big]^{2}\bigg\}. 
\label{eqn:min}
\end{equation}
\noindent
To find the minimum, we can take partial derivatives with respect to $\alpha$ and $\beta$, respectively, and solve the system of equations. The resulting minimizer leads to a credibility premium with past loss experience as
\begin{align}
P_{R}
=&
\widehat{\alpha}\,+\widehat{\beta} \, \widehat{R}({\bf{X}})\nonumber \\ =&\frac{\Cov\Big(\widehat{R}({\bf{X}}),\mu_{R}(\theta)\Big)}{\Var\Big(\widehat{R}({\bf{X}})\Big)}\, \widehat{R}({\bf{X}})+\Bigg[1-\frac{\Cov\Big(\widehat{R}({\bf{X}}),\mu_{R}(\theta)\Big)}{\Var\Big(\widehat{R}({\bf{X}})\Big)} \, 
\frac{\E\Big[\widehat{R}({\bf{X}})\Big]}{\mu}\Bigg] \, \mu.
\label{eqn:premium1}
\end{align}

Next, we investigate the asymptotic properties of 
this proposed premium with trimmed and winsorized data, respectively.\\

\subsection{Asymptotic Properties for Trimmed Data}

\noindent
The sample trimmed mean $\widehat{T}({\bf{X}})$ in equation (3.1) distributively converges to the population trimmed mean $\mu_{T}(\theta)$ in equation (\ref{eqn:moment_T}). Besides, \cite{Brazauskas2009} have shown that for each fixed $\theta$, the process variance of trimmed mean is
\begin{align}
v_{T}(\theta)
&=
\frac{1}{(1-p-q)^2}
\int_{p}^{1-q}\int_{p}^{1-q}
\left(
\text{min}\{u,v\}-uv
\right) \, 
dF^{-1}(u)\,dF^{-1}(v).
\label{eqn:var_T}
\end{align}
\noindent
Thus the derived robust estimator in the loss 
model has the following asymptotic normality
\begin{equation}
   \sqrt{n}\Big[\widehat{T}({\bf{X}})-\mu_{T}(\theta) \Big]\, \sim \, \mathcal{AN} \Big(0, v_{T}(\theta) \Big).
\end{equation}

\begin{note}
Compared with the variance derived through influence 
function by \cite{Kim2013}, equation (\ref{eqn:var_T}) 
has two advantages: 
(i) it can generally be applied to any distribution
in the scale-shape as well as location-scale
families of distributions and 
(ii) at least what we observed via simulation is 
that the variability is less volatile than the 
corresponding approximation by influence function. 
The theoretical aspect of the second fact
invites future topic to work on this 
area of research. 
\qed 
\end{note}

\subsection{Asymptotic Properties for Winsorized Data}

\noindent
By \cite{Serfling1980}, the sample winsorized mean in equation (3.1)  can be written as
\begin{align}
\widehat{W}({\bf{X}})=\frac{1}{n}\sum_{i=1}^{n}J \bigg(\frac{i}{n+1}\bigg)X_{(i)}+c_{n}^{(1)}X_{([np^{(1)}])}+c_{n}^{(2)}X_{([np^{(2)}])}
\label{eqn: sample_W}
\end{align}
where 
\begin{equation*}
J(x) = {\bf{1}}\bigg\{p^{(1)}\leq x\leq p^{(2)}\bigg \}=\begin{cases}
1, & \text{if } np^{(1)} \leq x \leq np^{(2)}; \\[12 pt]
0,& \text{otherwise}; \\
\end{cases}
\end{equation*}
with $p^{(1)}= p$ and $p^{(2)}=1-q$  and where $p$ and $q$ represent left and right winsorizing proportions, respectively. Also, $\displaystyle{\lim_{n\to \infty}c_n^{(1)} = c^{(1)}=p^{(1)}= p}$ and  $\displaystyle{\lim_{n\to \infty}c_n^{(2)} = c^{(2)}=p^{(2)}= 1-q}$. \\

\noindent
\cite{Chernoff1967}, have demonstrated that for each fixed $\theta$, 
\begin{equation}
    \widehat{W}({\bf{X}})\overset{d}{\longrightarrow} \mu_{W}(\theta),
    \label{eqn: mean_W}
\end{equation}
and 
\begin{align}
\Var\Big(\widehat{W}({\bf{X}})\Big) \overset{d}{\longrightarrow} v_{W}(\theta)=\int_{0}^{1}\alpha^{2}(u)du,
\label{eqn:var_W_def}
\end{align}
where 
\begin{align*}
\alpha(u)=\frac{1}{1-u}{\int_{u}^{1}J(w)H^{'}(w)(1-w)dw+\sum_{m=1}^{2} {\bf{1}}\bigg\{p^{(m)}\geq u\bigg\}c^{(m)}\Big(1-p^{(m)}\Big)H^{'}\Big(p^{(m)}\Big)}
\end{align*}
and $H(w)=F^{-1}(w)$.
Thus the derived robust estimator in the loss model has asymptotic normality
\begin{equation}
    \sqrt{n}\Big[\widehat{W}({\bf{X}})-\mu_{W}(\theta) \Big]\, \sim \, \mathcal{AN} \Big(0, v_{W}(\theta) \Big).
\end{equation}

\noindent
 \cite{zhao_brazauskas_ghorai_2018} have proved that equation (\ref{eqn:var_W_def}) is equivalent to
\begin{align}
 v_{W}(\theta)=&\,\Var(X_{W}|\theta)+2\Big[\mu_{W}(\theta) (A-B)+B H(1-q)-A  H(p)\Big] -(A-B)^{2}+\frac{A^2}{p}+\frac{B^2}{q},
\label{eqn:var_W}
\end{align}
\noindent
where $A = p^{2}H^{'}(p)$ and $B=q^{2}H^{'}(1-q)$. \\

\subsection{Robust Credibility Premium}

To build the linear credibility premium in
trimmed and winsorized data,
we also prove the following results. 

\begin{lemma}
Assume that $\mathbb{E}(|X_{i}|)<\infty$ and 
$\mathbb{E}(|X_{i}|^{2})<\infty$.
Then as $n\to \infty$, we have 
\begin{enumerate}[label=(\roman*)]
\item 
$\E\Big[\E\Big[\widehat{R}({\bf{X}})\Big]\Big]\to \mu_{R}$,

\item 
$n\E\Big[\Var\Big(\widehat{R}({\bf{X}})\Big)\Big] \to v_{R}$, 

\item 
$Var\Big(\E\Big[\widehat{R}({\bf{X}})\Big]\Big)\to a_{R}$

\item 
$\Cov\Big(\widehat{R}({\bf{X}}),\mu_{R}(\theta) \Big)\to a_{R}$, 

\item 
$ 
\Var\Big(\widehat{R}({\bf{X}})\Big)\to a_{R}+\dfrac{v_{R}}{n}.
$
\end{enumerate}

\begin{proof}
The properties of the trimmed mean have been 
investigated through \cite{Kim2013}. 
Here, we just prove for the  winsorized moment. 

\begin{enumerate}[label=(\roman*)]
\item 
By (\ref{eqn: mean_W}) $\E\Big[\widehat{W}({\bf{X}})\Big] \to \mu_{W}(\theta)$, then $\E\Big[\E\Big[\widehat{W}({\bf{X}})\Big]\Big] \to \E\Big[\mu_{W}(\theta)\Big]=\mu_{W}$ as $n\to \infty$.

\item 
By (\ref{eqn:var_W_def}), $n\E\Big[\Var\Big(\widehat{W}({\bf{X}})\Big)\Big] = \E\Big[nVar\Big(\widehat{W}({\bf{X}})\Big)\Big]\to \E\Big[v_{W}(\theta)\Big]=v_{W}$.

\item 
Again, 
by (\ref{eqn: mean_W}), 
$\Var\Big(\E\Big[\widehat{W}{(\bf{X}})\Big]\Big)\to \Var\Big(\mu_{W}(\theta)\Big)=a_{W}$.

\item 
The covariance can be written as 
\[
\Cov\Big(\widehat{W}({\bf{X}}),\mu_{W}(\theta) \Big)
=
\E\Big[\widehat{W}({\bf{X}}) \mu_{W}(\theta) \Big]-\E\Big[\widehat{W}({\bf{X})}\Big]\E\Big[\mu_{W}(\theta) \Big].
\]
Since $\Big|\widehat{W}({\bf{X}}) \mu_{W}(\theta)\Big|\leq \Big|\widehat{W}({\bf{X}})\Big| \Big|\mu_{W}(\theta)\Big|$, and both $\Big|\widehat{W}({\bf{X}})\Big|$ and $\Big|\mu_{W}(\theta)\Big|$ are square integrable, 
then by dominated convergence theorem
\citep[see, e.g.,][p. 54]{MR1681462}, 
\[
\Cov\Big(\widehat{W}({\bf{X}}),\mu_{W}(\theta) \Big)
\to \E\Big[\mu_{W}^{2}(\theta) \Big]-\Big(\E\big[\mu_{W}(\theta)\big]\Big)^{2}
=
a_{W}. 
\]

\item 
The law of total variance gives us
\begin{align*}
\Var\Big(\widehat{W}({\bf{X}})\Big)
& = 
\Var\Big(\E\Big[\widehat{W}({\bf{X}})\big|\theta\Big]\Big)+\E\Big[\Var\Big(\widehat{W}({\bf{X}})\big|\theta\Big)\Big] \\ 
& \to 
\Var\Big(\mu_{W}(\theta)\Big)+\E\Big[\Var\Big(\widehat{W}({\bf{X}})\Big)\Big]=a_{W}+\dfrac{v_{W}}{n}.
\end{align*}
Based on these asymptotic properties, the credibility premium from equation (\ref{eqn:premium1}) is given as
\begin{align}
P_{W} \quad
\to \quad & \frac{a_{W}}{a_{W}+\dfrac{v_{W}}{n}} \widehat{W}({\bf{X}})+\Bigg(1-\frac{a_{W}}{a_{W}+\dfrac{v_{W}}{n}} \frac{\mu_{W}}{\mu_{W}}\Bigg) \mu_{W} \nonumber\\[1.25ex]
& = Z_{W}  \widehat{W}({\bf{X}})+(1-Z_{W}) \mu_{W}, \label{eqn: credibility}
\end{align}
where 
$$Z_{W}=\dfrac{n}{n+v_{W}/a_{W}}.$$
\qedhere 
\end{enumerate}
\end{proof}
\end{lemma}

Next, we will see how this robust credibility premium works on the parametric models and non-parametric cases. \\

\section{Parametric Examples}
\label{sec:ParametricExamples}

In this section, we use two pairs of parametric combinations, to demonstrate the performance of data methodologies, trimming and winsorizing, towards the credibility premium estimations. 
In each comparison, the risk parameter $\theta$ follows the common parametric model. This setting can help us understand
how the loss likelihood model solely (the risk distribution is fixed) affects the ultimate premium estimation. In particular, we investigate the model stability and estimation consistency with various underlying model assumptions and robust schemes.\\

To display the general application of parametric models, we choose one comparison from the scale-shape distribution family, and the other comes from the location-scale family.

\subsection{Scale-shape Distribution Family Example}

The first pair of comparison is between the Exponential - Gamma Model and Pareto - Gamma Model.
Both Exponential and Pareto are heavy-skewed distributions that are typically used to fit the loss data. Pareto distribution resembles the shape of Exponential but has a heavier tail for extreme claim quantities. Now, we discuss their asymptotic properties under the proposed robust credibility structures and then compare their model performance under robust data trimming and data winsorizing, respectively.\\

Let $X_{1}|\theta, \hdots, X_{n}|\theta$ be i.i.d. random variables, following an Exponential distribution with a mean of $\theta$,
that is, 
$\theta$ is the scale parameter. And the parameter $\theta$ is  Gamma$(\alpha, \beta)$ distributed with mean of $\alpha/\beta$ and variance of $\alpha/\beta^{2}$.
Now the credibility premium for the Exponential - Gamma model is derived as follows. 

For loss random variable $X|\theta \sim \mbox{Exp}(\theta)$, we have $F(x|\theta)=1-e^{-x/\theta}=w$ and the quantile function $F^{-1}(w)=-\theta\text{log}(1-w)$. We will look at the structure formulas for trimmed mean and winsorized mean, separately.\\

\noindent
{\sc{Trimmed Version:}}\\
By (\ref{eqn:moment_T}), the hypothetical mean of the Exponential model is \begin{align*}
  \mu_{T}(\theta)&=
  -\theta\Big[\frac{1}{1-p-q}\int_{p}^{1-q}\text{log}(1-w)dw\Big]=\theta\,\dfrac{(1-p)[1-\text{log}(1-p)]-q(1-\log q)}{1-p-q} \\
  &:=\theta\,m_{1T}.
\end{align*}
The corresponding process variance is derived from (\ref{eqn:var_T}), that is
\begin{align*}
v_{T}(\theta)&=\theta^2\frac{1}{(1-p-q)^2}\int_{p}^{1-q}\int_{p}^{1-q}(\text{min}\{u,v\}-uv)\,\frac{1}{(1-u)}\,\frac{1}{(1-v)}\,du\,dv \\
& := \theta^2\,m_{3T}.
 \end{align*}
\noindent
{\sc{Winsorized Version:}}\\
By (\ref{eqn:moment_W}), the hypothetical mean of the Exponential distribution is
\begin{align*}
  \mu_{W}(\theta)&=
  -\theta\Big[p\text{log}(1-p)+\int_{p}^{1-q}\text{log}(1-w)dw+q\log q\Big]=\theta \Big[ 1-p-q-\text{log}(1-p)\Big], \\
  &:=\theta\,m_{1W}.
\end{align*}
Regarding the process variance, 
we start with the formula for winsorized variance, that is
\begin{align*}
\Var(X_{w}|\theta)
&=
\E[X_{w}^{2}|\theta]-\E^{2}[X_{w}|\theta] \\
& = 
\theta^{2}\bigg\{p\big[\text{log}(1-p)\big]^{2}+\int_{p}^{1-q}\big[\text{log}(1-w)\big]^{2}\,dw+q\big(\log q)^{2}\bigg\}-[\mu_{W}(\theta)]^{2} \nonumber\\
&=\theta^{2}\bigg\{ \underbrace{\big[\text{log}(1-p)\big]^{2}+2(1-p)\big[1-\text{log}(1-p)\big]-2q(1-\text{log}\, q)}_{m_{2W}} \bigg\}-\theta^{2}m_{1W}^{2}\nonumber\\
&:=\theta^{2}\Big(m_{2W}-m_{1W}^{2}\Big).
\end{align*}
\noindent
Therefore, the process variance by (\ref{eqn:var_W}) is
\begin{align*}
   v_{W}(\theta)&=\theta^{2}\bigg\{m_{2W}-m_{1W}^{2}+2\bigg[m_{1W}\bigg(\frac{p^{2}}{1-p}-q\bigg)-q\,\log q+\frac{p^{2}}{1-p}\,\text{log}(1-p)\bigg]\nonumber\\
   &\quad+\frac{p^{3}}{1-p}+q(1-q)+\frac{2p^{2}q}{1-p}\bigg \}\nonumber\\
   &:=\theta^{2} m_{3W}.
\end{align*}
See detailed derivation in appendix A.1.\\

Since above mentioned $m_{1}, m_{2}$, and $m_{3}$ only depend on the trimming or winsorizing proportions $p$ and $q$, not the risk parameter $\theta$, thus we treat them as constants in the following robust structural parameters.\\ 

\noindent
{\emph{Note}}: $m_{1}$ represents the general notation of $m_{1T}$ and $m_{1W}$, and the same for the other $m$ notations.\\

\noindent
With $\alpha>0$ and $\beta>0$, the collective premium, the variance of hypothetical mean, and the expectation of process variance for this Exponential - Gamma model are determined as
\begin{align}
\mu_{R}&=\E[\mu_{R}(\theta)]=\E\big[\theta m_{1}\big]=\E\big[\theta\big]m_{1}=\frac{\alpha}{\beta}m_{1},\\
a_{R}&=\Var(\mu_{R}(\theta))=\Var\big(\theta m_{1}\big)=\Var\big(\theta \big)\,m_{1}^{2}=
\frac{\alpha}{\beta^{2}}\,m_{1}^{2}, \nonumber   \\
v_{R}&=\E[v_{R}(\theta)]=\E\big[\theta^{2} m_{3}\Big]=\E\big[\theta^{2}\big]m_{3}=\dfrac{\alpha(\alpha+1)}{\beta^{2}}m_{3}.
\label{eqn:structure}
\end{align}
Finally, the credibility factor is
$$Z_{R}=\dfrac{n}{n+\frac{v_{R}}{a_{R}}}=\dfrac{n}{n+(\alpha+1)\frac{m_{3}}{m_{1}^{2}}}$$
and  by (\ref{eqn: credibility}) the credibility premium for this model is $$P_{R}=Z_{R}\, \widehat{R}({\bf{X}})+(1-Z_{R})\, \mu_{R}
,\qquad R\in (W, T).$$
When $(p,q)\to (0,0)$, the values  $m_{1}\to 1, m_{2}\to 2$, and $m_{3}\to 1$, both the trimmed mean and winsorized mean converge to the sample mean $\Bar{X}$. Thus
$$Z_{R}\to \dfrac{n}{n+(\alpha+1)}$$
and the credibility premium becomes
\[ \lim_{(p,q)\to (0,0)}P_{R}= \dfrac{n}{n+(\alpha+1)}\,\Bar{X}+\dfrac{\alpha+1}{n+(\alpha+1)}\,\mu, \]
where $\mu=\dfrac{\alpha}{\beta}$ is the collective premium.\\

For the compared Pareto-Gamma model, 
the loss random variable $X|\theta \sim \mbox{Pareto}(t,\theta)$, and
we have $F(x)=1-\left(\dfrac{\theta}{x+\theta}\right)^{t}=w$.
The robust structural parameters of credibility
premium can be derived through 
equations (\ref{eqn:moment_T}),
(\ref{eqn:moment_W}),
(\ref{eqn:var_T}), and 
(\ref{eqn:var_W}). This leads to the same structure credibility parameter as (\ref{eqn:structure}), 
where the $m_{1}, m_{2}$ and $m_{3}$ are general notations for both $T$ and $W$ parameters and detailed described in appendix A.2.\\

Hence, the credibility factor
$$Z_{R}=\dfrac{n}{n+(\alpha+1)\frac{m_{3}}{m_{1}^{2}}}.$$
When $(p,q)\to (0,0)$, $m_{1}\to \dfrac{1}{t-1}, m_{2}\to \dfrac{t}{t-2}- \dfrac{2t}{t-1}+1$, and $m_{3}\to \dfrac{t}{t-2}- \dfrac{2t}{t-1}+1-\left(\dfrac{1}{t-1}\right)^{2}$. Therefore
$$Z_{R}\to \dfrac{n}{n+(\alpha+1)\frac{t}{t-2}}$$
and the credibility premium for complete loss data is
\[ \lim_{(p,q)\to (0,0)}P_{R}= \dfrac{n}{n+(\alpha+1)\frac{t}{t-2}}\,\Bar{X}+\dfrac{(\alpha+1)\frac{t}{t-2}}{n+(\alpha+1)\frac{t}{t-2}}\,\mu. \]
Here, the collective premium has the format of $\mu=\dfrac{\alpha}{\beta(t-1)}$.

\subsection{Location-scale Family Distribution Example}

In contrast to the Exponential and Pareto distributions that belong to the shape-scale family, we pick up Lognormal-Normal and Log-logistic-Normal (the typical log-transformation of location-scale family distributions) in the second pair of comparisons and show the general application of the proposed approach.

Now we look at the detailed derivation of the Log-logistic-Normal. For loss random variable $X|\theta \sim \mbox{Log-logistic}(\theta, \sigma^2)$, and $\theta \sim \mbox{Normal}(\mu, v^2)$, we have the cdf of conditional distribution $$F(x|\theta)=\frac{1}{1+e^{-\frac{\log x-\theta}{\sigma}}},$$
where $-\infty<\theta<\infty$ and $\sigma<1$. This parameter setting guarantees the existence of the mean and variance.
And the quantile function $F^{-1}(w)=e^{\theta}\left(\dfrac{w}{1-w}\right)^{\sigma}$. 

\medskip 

\noindent
{\sc{Trimmed Version:}}\\
By equations (\ref{eqn:moment_T}) and (\ref{eqn:var_T}), the robust hypothetical mean and process variance are
\begin{align*}
\mu_{T}(\theta)&=\frac{e^{\theta}}{1-p-q}\int_{p}^{1-q} \left(\frac{w}{1-w}\right)^{\sigma}dw:=e^{\theta}\, m_{1T}, \\
v_{T}(\theta)
&=e^{2\theta}\frac{1}{(1-p-q)^2}\int_{p}^{1-q}\int_{p}^{1-q}(\text{min}\{u,v\}-uv)\,\sigma\frac{ u^{\sigma-1}}{(1-u)^{\sigma+1}}\,\sigma\frac{ v^{\sigma-1}}{(1-v)^{\sigma+1}}\,du\,dv\\
&:=
e^{2\theta}\, m_{3T}.
\end{align*}

\noindent
{\sc{Winsorized Version:}}\\
By equations (\ref{eqn:moment_W}) and (\ref{eqn:var_W}), the robust estimators are derived as
\begin{align*}
\mu_{W}(\theta)&=e^{\theta}\left\{p\left(\frac{p}{1-p}\right)^{\sigma}+\int_{p}^{1-p} \left(\frac{w}{1-w}\right)^{\sigma}dw+ q\left(\frac{1-q}{q}\right)^{\sigma}\right\}:=e^{\theta}\,m_{1W},\\
\Var(X_{w}|\theta)
&=
p \big[F^{-1}(p)\big]^{2}+ \int_{p}^{1-q}\big[F^{-1}(w)\big]^{2}\,dw+b\big[F^{-1}(1-q)\big]^{2}-[\mu_{W}(\theta)]^{2} \\
& = 
e^{2\theta}\big(m_{2W}-m_{1W}^{2}\big), \\
v_{W}(\theta)   
& = 
e^{2\theta} 
\Bigg\{m_{2W}-m_{1W}^{2} -\left(p^2 \Delta_{p}-q^2 \Delta_{q}\right)^2+p^3 \Delta_{p}^2 + q^3 \Delta_{1-q}^2\\
& \quad  
+2
\bigg[ 
m_{1W} (p^2 \Delta_{p}-q^2 \Delta_{1-q} )+q^2 \Delta_{1-q}\, 
\left(\frac{1-q}{q}\right)^{\sigma}
-p^2 \Delta_{p}\, \left(\frac{p}{1-p}\right)^{\sigma}
\bigg]
 \Bigg\}
\nonumber\\
&:=e^{2\theta}\,m_{3W},
\end{align*}
where
$
\Delta_{p}=\sigma\dfrac{ p^{\sigma-1}}{(1-p)^{\sigma+1}}$ and $\Delta_{1-q}=\sigma\dfrac{(1-q)^{\sigma-1}}{q^{\sigma+1}}
$.\\

Therefore, if the risk parameter $\theta$ is normally distributed with mean of $\mu$ and variance of $v^{2}$, the three structural parameters of credibility premium are
\begin{align}
\mu_{R}&=\E[e^{\theta}]\,  m_{1}=e^{\mu+\frac{1}{2}v^2}\, m_{1},
\nonumber\\ 
a_{R}&=\Var(e^{\theta})\, m_{1}^{2}=\Big(e^{2\mu+2v^2}-e^{2\mu+v^2}\Big)\,m_{1}^{2}, \nonumber\\
v_{R}&=\E[e^{2\theta}]\,m_{3}=e^{2\mu+2v^{2}}\,m_{3}.
\label{eqn:structure_LN}
\end{align}
Finally, the credibility factor is
\begin{equation}
 Z_{R}=\dfrac{n}{n+\frac{e^{2\mu+2v^{2}}m_{3}}{e^{2\mu+v^2}(e^{v^2}-1) m_{1}^{2}}}.
 \label{eqn: Z_Log}
\end{equation}
In this model, when $(p,q)\to (0,0)$, $m_{1}, m_{2}$, and $m_{3}$ depend on variability $\sigma$ and the integral of the derivative of quantile functions. The robust alternative means still converge to the sample mean $\Bar{X}$. \\

For the counterpart of Lognormal - Normal model, we have the loss random variable $X|\theta \sim LN(\theta, \sigma^{'2})$ and prior distribution $\theta \sim N(\mu, v^2)$. The cdf of loss likelihood is
$F(x|\theta)=\Phi\left(\dfrac{\ln x-\theta}{\sigma^{'}}\right)=w$ and the quantile function becomes $F^{-1}(w)=e^{\theta+\sigma^{'}\Phi^{-1}(w)}$.
The robust collective premium, variance of hypothetical mean, and expectation of process variance can be derived through equations (\ref{eqn:moment_T}),
(\ref{eqn:moment_W}),
(\ref{eqn:var_T}), and 
(\ref{eqn:var_W}). And the structural formulas in credibility model are exactly the same as in (\ref{eqn:structure_LN}) and (\ref{eqn: Z_Log}),
 but the $m_{1}, m_{2}$ and $m_{3}$ are described separately, in appendix A.3. In this combination, when $(p,q)\to (0,0)$, the values  $m_{1}\to e^{\frac{1}{2}\sigma^{'2}}, m_{2}\to e^{2\sigma^{'2}},$ and $m_{3}\to e^{2\sigma^{'2}}-e^{\sigma^{'2}}$.

\section{Simulation Study}
\label{sec:Sim}

Now we use the above-developed four credibility models
to illuminate the benefits of the proposed robust credibility premium estimation procedure and compare the performance of robust estimators based on trimming and winsorizing data, respectively.

\subsection{Study Design}

First, Monte Carlo simulation is used to approximate a continuous distribution of underlying risk level $\theta$. We randomly generate $n$ values of  $\theta$ from prior distribution $\pi(\theta)$, and these $\theta s$ keep the same for a fixed sample size $n$. Under each $\theta$, the conditional loss is determined by a contamination model
\begin{align}
\label{eqn:PPYZ_Mixture1}
F_{\epsilon}(x|\theta) 
& = 
(1-\epsilon) 
F_{0}(x|\theta) 
+ 
\epsilon\,
F_{c}(x|\theta),
\end{align}
where $\epsilon$ represents the probability that a sample 
observation comes from the contaminating distribution $F_{c}$ 
instead of the central (assumed) model $F_{0}$.
That is, $\epsilon = 0$ simply  corresponds to $F_{0}$.
The main objective is to see how estimated structure parameters (under $F_0$) behave when the actual data slightly deviates from the assumed model.\\

With generated $N$ data points under a specific risk level $\theta$, we apply both $T$ and $W$ robust procedures to take against outliers and then determine the structure parameters for credibility estimation in the overall portfolio. Various right trimming/winsorizing proportions $q$ and contamination rates $\epsilon$ are selected to investigate how the two robust approaches handle with model mis-specification. The simulation study is designed using the following setups:
\begin{itemize}
\item[(i)] Loss model and contamination combinations:\\ 
(a) $\theta\sim \text{Gamma}(4,2), F_{0}(x|\theta)\sim \text{Exponential}(\theta/2), F_{c}(x|\theta)\sim \text{Pareto}(\theta, 3)$;\\
(b) $\theta\sim \text{Normal}(4,1), F_{0}(x|\theta)\sim \text{Lognormal}(\theta, 0.45), F_{c}(x|\theta)\sim \text{Log-logistic}(\theta, 0.45)$.
\item[(ii)] Sample size: $n=1000$ and $N=100$.
\item[(iii)]  
Trimming and Winsorizing proportions: 
$p \equiv 0$ and
$q=0.00, 0.01, 0.05, 0.10, 0.20$.
\item[(iii)]  Contamination proportions: 
$\epsilon =0, 0.01, 0.03, 0.06, 0.10$.
\end{itemize}

The choice of model parameters, sample size, and trimming/winsorizing probabilities depends on practical considerations and research objectives. For example, 1000 samples of $\theta$ are enough to capture continuous prior for risk groups. And 100 losses for each $\theta$ are suitable to apply the trimming or winsorizing framework to robust credibility estimation. Also,  this size avoids the full credibility situation in which the credibility parameter $k=v/a$ slightly affects the calculation of credibility factor $z=\frac{N}{N+k}$.\\

More importantly, we want to investigate the effect of various contamination levels by controlling the parameters from two perspectives: 
\begin{itemize}
\item The heaviness of contaminating distribution:
In the first contamination, we deliberately choose $F_{0}\sim \text{Exponential}(\theta/2)$ and $ F_{c}\sim \text{Pareto}(\theta, 3)$ such that the two models have a common hypothetical mean $\mu(\theta)=\theta/2$ at $p=q=0$ (non-robust situation). And the risk parameter $\theta$
follows the same Gamma$(4,2)$, leading to an identical collective premium of $\mu = 1$, regardless of the model structure. Therefore, the similar behavior of these two loss distributions makes this combination a lighter contamination model. In the second contamination, the same location parameter $\theta$ also leads to the common hypothetical mean as $(p,q)\to (0,0)$. However, the same scale parameter $\sigma=0.45$ will significantly increase the collective premium based on the contaminated log-logistic losses, compared with that of the center lognormal model. Thus, the impact of heavier contamination can be observed.
\item The contamination proportion:  small values of  $\epsilon\, (0.01,0.03)$ help to indicate the impact of a single or few outliers in credibility estimation and large value of $\epsilon \,(0.06,0.10)$ illustrate how does robust and non-robust procures respond to the model mis-specification in credibility estimation.
\end{itemize}

We repeat the entire simulation 10 times and all computation is running through {\emph{R}}. 
The standardized ratio, 
e.g., $\mathbb{E} ( \hat{\mu} ) / \mu$,
that we report 
is defined as the average of 10  estimates divided by 
the true value of the structure parameter generated by the center model.  Using the uncontaminated model as the Benchmark, we compare the performance of each robust premium and discuss the corresponding bias that has been generated. 

\begin{table}[hbt!]  
\caption{Relative bias of structural parameters
$\widehat{\mu}, \
\widehat{v}, \ 
\widehat{a}$, and credibility parameter $\widehat{k}$ for Exponential-Gamma and Pareto-Gamma models at
selected contamination levels $\varepsilon$, with $n=1000$ and $N=100$. 
}
\label{table:ExpGamma1}
\vspace{-12pt}
\begin{center}
\begin{tabular}{l|c|ccccc|ccccc}
\hline
\multicolumn{1}{l}{Contam.} &
\multicolumn{1}{|c|}{Structure} &
\multicolumn{10}{c}{Robust Method} \\[-0.5ex]
\cline{3-12}
\multicolumn{1}{l}{Rate} &
\multicolumn{1}{|c|}{Parameter} &
\multicolumn{5}{c|}{Trimming with proportion $q$} &
\multicolumn{5}{c}{Winsorizing with proportion $q$}
\\[-0.5ex]
\cline{3-12}
 & & 0.00 & 0.01 & 0.05 & 0.10 & 0.20 
   & 0.00 & 0.01 & 0.05 & 0.10 & 0.20 \\[0.5ex]
\hline
\hline
\multirow{4}{*}{$\epsilon=0.00$} & $\frac{\mathbb{E} (\widehat{\mu})}{\mu}$  
& 1.01 & 1.00 & 0.96 & 0.91 & 0.81
& 1.01 & 0.97 & 0.95 & 0.93 & 0.89 \\

&  $\frac{\mathbb{E} (\widehat{v})}{v}$ 
& 1.01 & 1.03 & 0.84 & 0.71 & 0.54
& 1.01 & 0.98 & 0.97 & 0.95 & 0.95 \\

 & $\frac{\mathbb{E} (\widehat{a})}{a}$
& 1.03 & 0.94 & 0.74 & 0.58 & 0.37
& 1.03 & 1.01 & 0.93 & 0.84 & 0.66 \\

 & $\frac{\mathbb{E} (\widehat{k})}{k}$ 
& 0.98 & 1.10 & 1.14 & 1.22 & 1.45
& 0.98 & 0.96 & 1.04 & 1.14 & 1.45 \\

\hline
\multirow{4}{*}{$\epsilon=0.01$} & $\frac{\mathbb{E} (\widehat{\mu})}{\mu}$  
& 1.01 & 0.96 & 0.85 & 0.75 & 0.61
& 1.01 & 1.00 & 0.97 & 0.91 & 0.80 \\

&  $\frac{\mathbb{E} (\widehat{v})}{v}$ 
& 1.04 & 1.06 & 0.85 & 0.71 & 0.54
& 1.04 & 0.99 & 0.99 & 0.98 & 0.95 \\

 & $\frac{\mathbb{E} (\widehat{a})}{a}$
& 1.02 & 0.93 & 0.74 & 0.56 & 0.36
& 1.02 & 1.00 & 0.92 & 0.82 & 0.64 \\

 & $\frac{\mathbb{E} (\widehat{k})}{k}$ 
& 1.01 & 1.14 & 1.07 & 1.26 & 1.48
& 1.01 & 0.98 & 1.07 & 1.19 & 1.48 \\

\hline
\multirow{4}{*}{$\epsilon=0.03$} & $\frac{\mathbb{E} (\widehat{\mu})}{\mu}$  
& 1.01 & 0.96 & 0.85 & 0.75 & 0.60
& 1.01 & 1.00 & 0.95 & 0.90 & 0.80 \\

&  $\frac{\mathbb{E} (\widehat{v})}{v}$ 
& 1.09 & 1.11 & 0.85 & 0.75 & 0.53
& 1.09 & 1.01 & 0.98 & 0.98 & 0.95 \\

& $\frac{\mathbb{E} (\widehat{a})}{a}$
& 1.04 & 0.94 & 0.73 & 0.57 & 0.36
& 1.04 & 1.01 & 0.93 & 0.83 & 0.65 \\

& $\frac{\mathbb{E} (\widehat{k})}{k}$ 
& 1.06 & 1.19 & 1.17 & 1.27 & 1.48
& 1.06 & 0.99 & 1.06 & 1.19 & 1.46 \\
 
\hline
\multirow{4}{*}{$\epsilon=0.06$} & $\frac{\mathbb{E} (\widehat{\mu})}{\mu}$  
& 1.01 & 0.96 & 0.85 & 0.75 & 0.60
& 1.01 & 0.99 & 0.95 & 0.90 & 0.80 \\

&  $\frac{\mathbb{E} (\widehat{v})}{v}$ 
& 1.17 & 1.19 & 0.86 & 0.71 & 0.53
& 1.17 & 1.02 & 1.00 & 0.99 & 0.95 \\

 & $\frac{\mathbb{E} (\widehat{a})}{a}$
& 1.03 & 0.92 & 0.71 & 0.56 & 0.35
& 1.03 & 1.00 & 0.91 & 0.81 & 0.64 \\

 & $\frac{\mathbb{E} (\widehat{k})}{k}$ 
& 1.14 & 1.30 & 1.19 & 1.28 & 1.51
& 1.14 & 1.02 & 1.09 & 1.22 & 1.49 \\

 \hline
\multirow{4}{*}{$\epsilon=0.10$} & $\frac{\mathbb{E} (\widehat{\mu})}{\mu}$  
& 1.01 & 0.96 & 0.84 & 0.74 & 0.59
& 1.01 & 0.99 & 0.95 & 0.90 & 0.79 \\

&  $\frac{\mathbb{E} (\widehat{v})}{v}$ 
& 1.23 & 1.26 & 0.86 & 0.71 & 0.53
& 1.23 & 1.05 & 1.01 & 1.00 & 0.97 \\

& $\frac{\mathbb{E} (\widehat{a})}{a}$
& 1.03 & 0.92 & 0.71 & 0.55 & 0.35
& 1.03 & 1.00 & 0.91 & 0.81 & 0.63 \\

 & $\frac{\mathbb{E} (\widehat{k})}{k}$ 
& 1.20 & 1.37 & 1.21 & 1.29 & 1.53
& 1.20 & 1.05 & 1.10 & 1.24 & 1.54 \\
 \hline
\end{tabular}
\end{center}

\vspace{-0.20cm}

{\footnotesize {\sc Note}: 
The entries are the average 
of $\widehat{\mu}/\mu, \widehat{v}/v, \widehat{a}/a$ and $\widehat{k}/k$ 
computed using 
1,000,000 simulated values.
Standard errors of these entries are
$\leq 0.11$.}
\end{table} 

\begin{table}[hbt!]
\caption{Relative bias of structure parameters $\widehat{\mu}, \
\widehat{v}, \ 
\widehat{a}$, and credibility parameter $\widehat{k}$ for Lognormal-Normal and log-logistic-Normal models at
selected contamination levels $\varepsilon$, with $n=1000$ and $N=100$.}
\label{table:LogNormalNormal1}
\vspace{-12pt} 
\begin{center}
\begin{tabular}{l|c|ccccc|ccccc}
\hline
\multicolumn{1}{l}{Contam.} &
\multicolumn{1}{|c|}{Structure} &
\multicolumn{10}{c}{Robust Method} \\[-0.5ex]
\cline{3-12}
\multicolumn{1}{l}{Rate} &
\multicolumn{1}{|c|}{Parameter} &
\multicolumn{5}{c|}{Trimming with proportion $q$} &
\multicolumn{5}{c}{Winsorizing with proportion $q$}
\\[-0.5ex]
\cline{3-12}
 & & 0.00 & 0.01 & 0.05 & 0.10 & 0.20 
   & 0.00 & 0.01 & 0.05 & 0.10 & 0.20 \\[0.5ex]
\hline
\hline
\multirow{4}{*}{$\epsilon=0.00$} & $\frac{\mathbb{E} (\widehat{\mu})}{\mu}$  
& 0.97 & 0.96 & 0.91 & 0.87 & 0.80
& 0.97 & 0.97 & 0.95 & 0.93 & 0.89 \\

&  $\frac{\mathbb{E} (\widehat{v})}{v}$ 
& 0.92 & 0.94 & 0.80 & 0.71 & 0.62
& 0.92 & 0.88 & 0.87 & 0.84 & 0.87 \\

 & $\frac{\mathbb{E} (\widehat{a})}{a}$
& 0.92 & 0.88 & 0.80 & 0.72 & 0.62
& 0.92 & 0.91 & 0.88 & 0.84 & 0.77 \\

 & $\frac{\mathbb{E} (\widehat{k})}{k}$ 
& 1.00 & 1.06 & 1.00 & 0.98 & 1.01
& 1.00 & 0.97 & 1.00 & 0.99 & 1.12 \\

\hline
\multirow{4}{*}{$\epsilon=0.01$} & $\frac{\mathbb{E} (\widehat{\mu})}{\mu}$  
& 0.98 & 0.96 & 0.91 & 0.86 & 0.80
& 0.98 & 0.97 & 0.95 & 0.93 & 0.89 \\

&  $\frac{\mathbb{E} (\widehat{v})}{v}$ 
& 1.01 & 1.03 & 0.82 & 0.73 & 0.63
& 1.01 & 0.93 & 0.90 & 0.90 & 0.87 \\

 & $\frac{\mathbb{E} (\widehat{a})}{a}$
& 0.92 & 0.89 & 0.80 & 0.72 & 0.61
& 0.92 & 0.91 & 0.88 & 0.84 & 0.77 \\

 & $\frac{\mathbb{E} (\widehat{k})}{k}$ 
& 1.10 & 1.17 & 1.03 & 1.01 & 1.03
& 1.10 & 1.02 & 1.02 & 1.07 & 1.13 \\

\hline
\multirow{4}{*}{$\epsilon=0.03$} & $\frac{\mathbb{E} (\widehat{\mu})}{\mu}$  
& 0.98 & 0.96 & 0.91 & 0.87 & 0.80
& 0.98 & 0.98 & 0.96 & 0.94 & 0.89 \\

&  $\frac{\mathbb{E} (\widehat{v})}{v}$ 
& 1.26 & 1.29 & 0.84 & 0.75 & 0.65
& 1.26 & 1.01 & 0.96 & 0.92 & 0.95 \\

 & $\frac{\mathbb{E} (\widehat{a})}{a}$
& 0.94 & 0.89 & 0.80 & 0.73 & 0.61
& 0.94 & 0.92 & 0.89 & 0.85 & 0.77 \\

 & $\frac{\mathbb{E} (\widehat{k})}{k}$ 
& 1.34 & 1.44 & 1.05 & 1.04 & 1.06
& 1.34 & 1.09 & 1.09 & 1.08 & 1.22 \\
 
\hline
\multirow{4}{*}{$\epsilon=0.06$} & $\frac{\mathbb{E} (\widehat{\mu})}{\mu}$  
& 0.99 & 0.96 & 0.91 & 0.86 & 0.79
& 0.99 & 0.98 & 0.96 & 0.94 & 0.89 \\

&  $\frac{\mathbb{E} (\widehat{v})}{v}$ 
& 14.68 & 14.98 & 0.90 & 0.79 & 0.67
& 14.68 & 1.21 & 1.01 & 0.98 & 0.90 \\

 & $\frac{\mathbb{E} (\widehat{a})}{a}$
& 0.98 & 0.86 & 0.81 & 0.73 & 0.61
& 0.98 & 0.94 & 0.90 & 0.86 & 0.78 \\

 & $\frac{\mathbb{E} (\widehat{k})}{k}$ 
& 13.00 & 32.97 & 1.11 & 1.08 & 1.08
& 13.00 & 1.28 & 1.13 & 1.15 & 1.15 \\

 \hline
\multirow{4}{*}{$\epsilon=0.10$} & $\frac{\mathbb{E} (\widehat{\mu})}{\mu}$  
& 1.00 & 0.97 & 0.91 & 0.87 & 0.79
& 1.00 & 0.99 & 0.96 & 0.94 & 0.89 \\

&  $\frac{\mathbb{E} (\widehat{v})}{v}$ 
& 1.76 & 1.80 & 0.93 & 0.81 & 0.68
& 1.76 & 1.22 & 1.02 & 1.01 & 1.02 \\

& $\frac{\mathbb{E} (\widehat{a})}{a}$
& 0.97 & 0.91 & 0.81 & 0.73 & 0.61
& 0.97 & 0.94 & 0.90 & 0.86 & 0.78 \\

 & $\frac{\mathbb{E} (\widehat{k})}{k}$ 
& 1.83 & 1.98 & 1.15 & 1.11 & 1.11
& 1.83 & 1.30 & 1.13 & 1.19 & 1.32 \\
 \hline
\end{tabular}
\end{center}

\vspace{-0.20cm}
{\footnotesize {\sc Note}: 
The entries are the average 
of $\widehat{\mu}/\mu, \widehat{v}/v, \widehat{a}/a$ and $\widehat{k}/k$ computed using 
1,000,000 simulated values.} 
\end{table}

\subsection{Simulation Result}
We conduct two simulations and record the bias of collective premium $(\mu)$, expectation of process $v$, variance of hypothetical mean $a$, and credibility parameter $k$, respectively. For each ratio, the closer the value is to 1, the better the estimation performs. \\

{\sc Discussion of Table \ref{table:ExpGamma1}:}
The contaminated model has the similar density compared to the base model.
As expected, a small right $T/W$-proportion \(q\) aligns 
the robust estimated ratios more closely with the non-robust theoretical values. 
As we transition from the top row to the bottom row, 
we observe a clear increase in bias as the contamination proportion increases.\\

By observing the first three ratios, namely 
\(\mathbb{E}(\widehat{\mu})/\mu\), 
\(\mathbb{E}(\widehat{v})/v\), and 
\(\mathbb{E}(\widehat{a})/a\), 
it is evident that the \(W\)-estimated 
values outperform the corresponding non-robust values
as the contamination proportion increases and 
the \(W\) proportion is reasonably chosen, 
being approximately aligned with the contamination level.
In other words, 
if the value of \(\epsilon\) is large and the \(W\) proportion 
is smaller but close to \(\epsilon\), 
the \(W\)-estimated values tend to outperform 
the corresponding non-robust values.
Additionally, 
when comparing \(T\)- and \(W\)-estimated ratios, 
\(W\) consistently outperforms when \(\epsilon > 0\).\\

In terms of ratio 
$\E(\widehat{k})/k$,
when examining the columns, 
it is noted that if the trimming/winsorizing proportions are close to the contamination proportion, 
then the robust performance metrics are weaker than their non-robust counterparts. 
However, 
the robust-winsorizing metrics are significantly superior
compared to the corresponding robust-trimming metrics. 
For example, with \(\epsilon = 0.06\) and \(q = 0.05\),
the ratios \(\mathbb{E}( \widehat{k} )/k\) 
for trimming and winsorization are 1.19 and 1.09, respectively, 
demonstrating a 10\% improvement in going from \(W\) to \(T\). 
Another critical observation is that, 
for any combination of contamination \(\epsilon\) and 
the right trimming/winsorizing proportions \(q\), 
all \(W\)-entries perform better than the corresponding \(T\)-entries,
except in the very last row and column.
This suggests that when the contaminated model 
has a density comparable to the base model, 
the robust-winsorized B{\"u}hlmann credibility parameter $\widehat{k}$ is more stable than the robust-trimmed version,
leading to a more stable credibility factor
\(Z = \frac{N}{N + \widehat{k}}\).
Of course, if there is no contamination, the non-robust ratios are closer to one.
However, it is important to note that as the contamination proportion, 
\(\epsilon\), increases, 
the winsorized ratios are closer to one compared 
to the corresponding non-robust ratios. 
For example, for \(\epsilon = 10\%\) and \(q = 5\%\),
the ratio \(\mathbb{E} ( \widehat{k} )/k\) is 1.10,
whereas the corresponding non-robust ratio is 1.20.
Therefore, for the B{\"u}hlmann credibility parameter \(\widehat{k}\), 
it is recommended to use the \(W\)-approach as a robust alternative 
compared to the corresponding \(T\) version, 
particularly if the tail behavior of the central 
model and the contaminated model is similar.
\\

{\sc Discussion of Table \ref{table:LogNormalNormal1}:}
The discussion of this table is very similar to Table \ref{table:ExpGamma1}; 
however, in this case, 
the contaminated model has a heavier density compared to the base model,
which leads to some important observations. 
Here, we present the discussion only for 
the ratio \(\mathbb{E} ( \widehat{k} )/k\). 
Since the contaminated model has a heavier tail, 
it is observed that higher trimming/winsorizing proportions yield better ratios.
For example, for \(\epsilon = 6\%\) and \(q = 5\%\), 
the \(T\)- and \(W\)-estimated ratios 
for \(\mathbb{E} ( \widehat{k} )/k\) are 1.11 and 1.13, respectively, 
whereas the corresponding non-robust ratio is 13.00, 
a significantly higher number. 
This clearly demonstrates the power of robust \(T/W\)-credibility models, 
given that the sample data comes from a contaminated model with a heavier 
tail compared to the base model. 
Another important observation from this table is that 
the higher the contamination- and lower the right \(T/W\)-proportions,
which we can call an under-robust situation,
the \(T/W\)-robust ratios \(\mathbb{E} ( \widehat{k} )/k\) 
are not better than the non-robust corresponding ratios.
However, as the \(T/W\)-proportions increase,
the \(T/W\)-robust ratios \(\mathbb{E} ( \widehat{k} ) /k\) 
improve compared to the non-robust ratios. 
Specifically, in this model with 
heavier contaminated model, 
we observed that the non-robust $(q=0)$ ratios 
have a standard error up to 0.80 and under-robust $q=0.01$ 
ratios have a standard error up to only 0.40.
Moreover, the higher the \(T/W\)-proportions, 
the better the \(T\)-performance compared 
to the corresponding \(W\)-performance. 
This suggests that over-robustifying 
essentially leads to less stable estimators.\\

An important and common observation from both 
Tables \ref{table:ExpGamma1} and \ref{table:LogNormalNormal1} 
is that the {\em B\"{u}hlmann credibility parameter} 
\(\widehat{k}\) 
exhibits a convex shape in terms of the right trimming/winsorizing  
proportion $q$ across both models and for both $T/W$ methods.
From the convexity of the \(\widehat{k}\) values, 
we can infer that over- or under-trimming/winsorizing  does 
not necessarily lead to lower or higher credibility. 
Instead, maximum credibility 
(i.e., minimum \(\widehat{k}\) value)
can be achieved with appropriate right 
trimming/winsorizing  proportions. 
For example, in Table \ref{table:LogNormalNormal1}, 
considering the case where 
\(\epsilon = 10\%\) and \(q = 5\%\), 
the minimum value for
\(\mathbb{E} ( \widehat{k} )/k = 1.13\) 
is observed for \(W\)-estimated ratios.
In other words, 
a small winsorizing proportion (compared to \(\epsilon\)) 
can maintain the stable performance of the structural 
parameter estimation while also addressing outliers.

\section{Non-parametric Estimation with Example}
\label{sec:NonPar}

\subsection{Non-parametric Procedure}

Empirical Bayes Credibility Models have been widely used in actuarial loss estimation (see \cite{Norberg1980} and \cite{Jin2017}).
Now we examine the empirical Bayes methods based on the robust trimmed or winsorized data (instead of original data), and display how to use the sample data to estimate the  $\mu_{R}, a_{R}$ and $v_{R}$ needed for building the credibility premium if the underlying parametric model is unavailable.\\

We illustrate the ideas with a simple Bühlmann-type model and we consider individual claim losses $(X_{ij})$ in the structure. Suppose there are $r (r\geq 1)$ groups of risks in a portfolio and the number of observations $n_{i}$ varies between risks. Then for $ith$ group, we have the risk parameter  $\theta_{i}$, and the loss vector
${\bf{X}}_{i}=(X_{i1},\hdots,X_{in_{i}}),\, i=1,\hdots,r.$\\

It is assumed that for each group $i$, $X_{ij}$ share the identical risk parameter $\theta_{i}$ and it does not change over time. Then the hypothetical mean $\mathbb{E}[X_{ij}|\theta_{i}]=\mu(\theta_{i})$ and process variance $\mathbb{V}ar(X_{ij}|\theta_{i})=v(\theta_{i})$.
Under each risk parameter $\theta_{i}$, the conditional $X_{ij}|\theta_{i}$ are assumed to be independent and identically distributed.  Our objective is to estimate the risk premium for each group $i$, using the robust credibility approaches. And then determine the total premium for the entire portfolio. \\

Let $R_{ij}$ be the trimmed or winsorized version of $X_{ij}$ by (\ref{eqn: T} or \ref{eqn: W}).
Then as the number of observations $n_{i}\to \infty$, the empirical hypothetical mean converges to the parameters such that
\begin{align}
&  \widehat{\mu}_{R}^{(i)}=\dfrac{\sum_{j=1}^{n_{i}}R_{ij}}{n_{i}^{'}} \to \mu_{R}(\theta_{i})
\label{eqn: nonpar_HM}
\end{align}
where $n_{i}^{'}$ is the number of observations after the trimming or winsorizing.
Besides, the empirical estimate, $\widehat{v}_{R}^{(i)}$, 
of process variance $v_{R}(\theta_{i})$ are empirical estimate of (\ref{eqn:var_T}) 
and (\ref{eqn:var_W}), respectively (see appendix A.4). All these lead to the robust version of empirical structural parameters 
\begin{align}
&\text{Collective Premium}
\quad &&
\widehat{\mu}_{R}=\frac{\sum_{i=1}^{r}n_{i}^{'} \, \widehat{\mu}_{R}^{(i)}}{\sum_{i=1}^{r}n_{i}^{'}}. \\
&\text{Expectation of Process Variance}\quad
&&
\widehat{v}_{R}=\frac{\sum_{i=1}^{r}n_{i}^{'}\, \widehat{v}_{R}^{(i)}}{\sum_{i=1}^{r}(n_{i}^{'}-1)}.
\label{eqn:EPV1} \\
\nonumber \\
&\text{Variance of Hypothetical Mean}\quad 
&&\widehat{a}_{R}
=
\frac{\sum_{i=1}^{r}n_{i}^{'}\, 
\big(\widehat{\mu}_{R}^{(i)}
-\widehat{\mu}_{R}\big)^{2}-(r-1)\widehat{v}_{R}}{N-\frac{1}{N}
\sum_{i=1}^{r}n_{i}^{'2}},
\quad 
N := \sum_{i=1}^{r} n_{i}^{'}.
\label{eqn:VHM1}
\end{align}

\begin{note}
Because we consider the individual loss instead of unit exposure loss or average loss in the setting, and the variance estimate is based on the theoretical asymptotic property, so the structure formulas are in a different format from the typical Bühlmann model (see \cite{Klugman2019}, page 448). 
\end{note}

Again, as the exposure units $n_{i}\to \infty$, these empirical estimates converge to the  parameters $\mu_{R}, a_{R}$ and $v_{R}$, respectively. 
Therefore, the credibility factor for each group $i$ is
$$Z_{R}^{(i)}=\frac{n_{i}}{n_{i}+\widehat{v}_{R}/\widehat{a}_{R}}$$
and the credibility premium for Group $i$ is estimated by
\begin{align}
\widehat{P}_{R}^{(i)}
&=
Z_{R}^{(i)}\,\widehat{\mu}_{R}^{(i)}
+
\left(1-Z_{R}^{(i)}\right)\,\widehat{\mu}_{R}.
\label{eqn: premium}
\end{align}

\subsection{Real Data Illustration}
\label{sec:RealData}

Next, a real data analysis is conducted to indicate the benefits of the proposed robust credibility premium estimations. The target under consideration is the Local Government Property Insurance Fund (LGPIF), an insurance pool administered by the Wisconsin Office of the Insurance Commissioner, which has been studied by \cite{Frees2018}. This insurance fund covers local government properties that include counties, cities, towns, villages, school districts, and library boards. Our goal is to investigate what effect initial assumptions have on the structure parameter estimation and the corresponding group credibility premiums in this insurance portfolio.\\

\begin{table}[htb]
\caption{Summary statistics of the loss data in various types of insurance property fund.}
\label{tab: summary}
\centering
\begin{tabular}{l|l|l|l|l|l}
\hline
Type  & Size & Min & Mean & Median & Max\\
\hline
City &329  & 501 & 17,368&	5,581	& 536,060 \\
County &359& 1,053&42,767&	25,659& 2,648,168\\
Library &34 & 700 &80,031&	5,917& 1,323,753	\\
School&486 &800&60,164&	25,202& 12,927,218\\
Town&28&1,250&6,655&	3,348& 71,101\\
Village&141&700&9,604& 4,000& 209,312\\
\hline
\end{tabular}
\end{table}

1377 LGPIF loss observations were recorded in 2010, and the summary statistics of each type of property are listed in Table \ref{tab: summary}.
It is clear that the loss data of each property type resembles a right heavy-tailed distribution. Thus, we will see how the upper trimming/winsorizing proportion $q$ can take against the extreme claims and affect the credibility premium ultimately. 
We set $q=(0,0.005, 0.01,0.02,0.05,0.10)$ in robust trimming and winsorizing, respectively, and then derive the structural parameters for  above-mentioned property funds through the procedure (\ref{eqn: nonpar_HM}) -- (\ref{eqn: premium}). The collective premium for each class and the total estimated premiums are displayed in Table \ref{tab:premium}.\\

\begin{table}[htb]
\caption{Collective premium (per claim) of each type of insurance property fund and the final estimated premium with trimmed ($T$) and winsorized ($W$) robust estimators.}
\label{tab:premium}
\vspace{-12pt}
{\footnotesize
\begin{center}
\begin{tabular}{r|l|l|l|l|l|l}
\hline
Type  & $q=0$ & $q=0.005$ & $q=0.01$ & $q=0.02$ & $q=0.05$ & $q=0.10$\\
\hline
City ($T$) &39,629& 19,546& 13,895& 12,186& 10,197 &8,637\\
($W$) & 39,629&19,485(-0.3)&15,789 (+13.6)& 14,184 (+16.4) &11,502 (+12.8)&10,644 (+23.2) \\

County ($T$) & 
\multirow{10}{*}{$\big\downarrow$} & 32,107& 32,309& 30,743 & 28,052 & 25,786\\
($W$)& &35,850 (+11.7)& 33,685 (+4.3) &32,700 (+6.4)& 31,194 (+11.2)& 29,313 (+13.7)\\

Library ($T$)& & 33,918 & 63,216 & {\bf 65,916}& 33,057& 23,191	\\
($W$)& & 43,209 (+27.4)& 64,984  (+1.2) &{\bf 61,834 (-6.2)}& 46,500 (+36.9)& 40,795 (+75.9)\\

School ($T$)& & 27,161& 24,734& 22,540& 19,679& 18,283\\
($W$)& & 31,405 (+15.6)& 26,940 (+8.9)& 25,260 (+12.1)& 21,850 (+11.0)& 19,942 (9.1)\\

Town ($T$)& & 23,472 & 12,347 & 10,881 & {\bf 9,896 } &6,053\\
($W$)& & 22,881 (-2.5)& 12,813 (+3.7)&12,974 (+19.2) &{\bf 8,074 (-18.4)} &7,479 (+23.6)\\

Village ($T$)& & {\bf 19,056} &9,589& 8,440& 7,200& 5,219\\
($W$)& & {\bf 16,578 (-13.0)}& 10,395 (+8.4) &10,260 (+21.6) &7,730 (+7.4)& 6,952 (+33.2)\\
\hline
Total ($T$) &54,568,809 &35,654,881 &32,037,976 &29,736,303 &\boxed{25,436,492}& 22,678,121\\
($W$)& 54,568,809 &38,990,823 (+9.4) &34,380,191 (+7.3)&32,594,918 (9.6) &28,498,888 (+12.0)&\boxed{26,293,544} (+15.9)\\
\hline
\end{tabular}
\end{center}

\vspace{-0.20cm}

{\sc Note:}
The value in parenthesis is the percentage change 
of premium from $T$ to $W$ estimation.
}
\end{table}

As is seen in Table \ref{tab:premium}, the impact of robust methods on the overall premium (per claim) estimation is consistent, all the winsorized total premiums have a higher level than that of the trimmed counterparts. However, this pattern was not shown for individual type of fund, which means, not every premium based on trimming is below the one based on winsorizing. Sample size seems to play an important role here. For a midsize sample data of Village, 0.5\% ($q=0.005$) of winsorizing will alternate the premium from 39,629 to 16,578, but the new estimation was  13\% less than the change with the trimming framework. Similar findings also have been displayed for the individual group of Town and Library.\\

Besides, it is clear to notice that
all types of losses yield to the same overall premium $39,639$ when $p=q=0$. Under such non-robust structure, the estimated variance of hypothetical mean $\hat{a}<0$, leads to an unreliable credibility estimation.
Another interesting finding is Library claims. Its premium estimation of Winsrozing credibility deviates tremendously from that of Trimming alternative. Small sample size ($n=34$) may amplify the effect of procedure, resulting in a 76\% premium increase when the largest 10\% values have been changed to robust version.\\

The actual collected premium of LGPIF is 25 million each year during 2006 - 2010. Obviously, the Buhlmann credibility premium with original loss leads to an estimation that more than double the real collected value. Meanwhile, 
25 million is close to the total estimated premium with 5\% of trimming and 10\% of winsorizing, respectively. The finding is very critical when the deductibles of the incoming policies are not known.

\section{Conclusion and Discussion}
\label{sec:Conclusion}
In this paper, 
we develop a robust B{\"u}hlmann credibility model based on
winsorizing the original sample data to mitigate the 
impact of outliers or model mis-specification 
in credibility premium estimation.
We examine the asymptotic normality properties
of the estimated credibility structural parameters 
for a variety of parametric models typically 
used for pricing insurance risks, 
following the general \(L\)-statistics approach.
The simulation study is conducted by varying the left and right winsorizing proportions from 0\% to 100\% for two competing conditional loss distributions, in which one resembles the shape of the other, but has a heavier tail for extreme values. Extending the 
theorem and outcome of \cite{zhao_brazauskas_ghorai_2018}, we also derive the non-parametric estimation procedure and analyze the sensitivity of winsorizing proportion to the credibility factor and ultimate premium estimation in the group insurance contract. To distinguish the impact of risk measures,  
all the results from the proposed winsorized scheme ($W$) are compared via the counterpart trimmed experience ($T$).\\

Finally, we discuss 
the major findings of this scholarly work, which include:
\begin{itemize}
\item Compared to the classical credibility approach, the use of robust credibility (both $T$ and $W$ cases) offers the advantage of preventing the effect caused by extreme losses or model mis-specifications. It can better capture the heavy tail of the underlying loss
models and thus improve the perspective of risk control to the insureds.
\item 
All the 
structural parameters via winsorizing are less volatile
compared to the corresponding quantities via trimming. In location-scale examples, the
winsorized scheme even could reduce the influence of model assumptions on credibility estimation and provide a more stable estimation for potential risk management.
\item A small proportion of winsorizing or trimming could significantly adjust the collective premium of a group insurance contract and avoid over-pricing issues.
And the $W$ procedure allows a larger proportion to handle extreme claims than the $T$ scheme, leading to more financially conservative insurers.
\end{itemize}

Our comprehensive study of robust credibility addresses the issue of over-estimated premiums due to extreme losses. The proposed methodology could apply to general parametric loss models and display an explicit structure in computation approximation, but there are also areas where some limitations and potential avenues for further research can be discussed. The current methodology is 
under a framework of pure robust credibility, a relation focusing on the robust estimate itself. \cite{PITSELIS2008} pointed out this framework may lead to a bias in the premium estimates.   \cite{Gisler1980} and 
\cite{Garrido2000} introduced an unbiased structure -- semi-credibility, but their central idea is still based on the implicit and volatile influence functions. To incorporate these limitations, we plan to enrich our work to provide a more comprehensive understanding of 
the relationship between the robust estimate and the non-robust estimate, develop the explicit format for an unbiased robust portfolio, and enhance the practical application of such credibility estimation. 
An additional avenue for potential future research
involves the exploration of implementing 
MTM/MWM approaches within the frameworks
of the Exponential or Esscher premium principles. 
This exploration would aim to incorporate the 
insurer's risk aversion more explicitly and effectively.

\newpage

\section*{Appendix}

\subsection*{A.1 Exponential-Gamma}
The integral of the second moment quantile function of the exponential is
\begin{align*}
 &\int_{p}^{1-q}\big[\log (1-w)\big]^{2}\,dw=-\int_{1-p}^{q}\big(\log 
 u\big)^{2}\,du=-\big[u(\log 
 u)^{2}+2u(1-\log u)\big]\big|_{1-p}^{q}\\&\hspace{1.5in}=(1-p)\big[\log (1-p)\big]^{2}+2(1-p)[1-\log (1-p)]-q(\log q)^{2}-2q(1-\log q).
\end{align*}
Thus the variance of winsorized mean is 
\begin{align*}
\Var(X_{w}|\theta)&= \E[X_{w}^{2}|\theta]-E^{2}[X_{w}|\theta]\nonumber \\
   &= p \big[F^{-1}(p)\big]^{2}+ \int_{p}^{1-q}\big[F^{-1}(w)\big]^{2}\,dw+q\big[F^{-1}(1-q)\big]^{2}-[\mu_{W}(\theta)]^{2}\nonumber \\
&=\dfrac{1}{\theta^{2}}\bigg\{p\big[\text{log}(1-p)\big]^{2}+\int_{p}^{1-q}\big[\text{log}(1-w)\big]^{2}\,dw+q\big(\log q)^{2}\bigg\}-[\mu_{W}(\theta)]^{2} \nonumber\\
  &=\dfrac{1}{\theta^{2}}\bigg\{p\big[\text{log}(1-p)\big]^{2}+(1-p)\big[\text{log}(1-p)\big]^{2}+2(1-p)[1-\text{log}(1-p)]-q(\text{log}\,q)^{2}\nonumber\\
  &\quad -2q(1-\text{log}q)+q\big(\text{log}\,q)^{2}\}-[\mu_{W}(\theta)]^{2}\nonumber\\
  &=\dfrac{1}{\theta^{2}}\bigg\{ \big[\log(1-p)\big]^{2}+2(1-p)\big[1-\log(1-p)\big]-2q(1-\log q) \bigg\}-\dfrac{1}{\theta^{2}}\,[\mu_{W}(\theta)]^{2}\nonumber\\
  &:=\dfrac{1}{\theta^{2}}\Big(m_{2W}-m_{1W}^{2}\Big).
\end{align*}
In equation (\ref{eqn:var_W}), the derivative of exponential quantile functions is
\begin{align*}
&H^{'}(w)=(F^{-1})^{'}(w)=\dfrac{1}{F^{'}(F^{-1}(w))}=\dfrac{1}{F^{'}\big(-\frac{\text{log}(1-w)}{\theta}\big)}=\dfrac{1}{\theta e^{-\theta \frac{-\text{log}(1-w)}{\theta}}}=\dfrac{1}{\theta(1-w)}.
\end{align*}

\noindent
Therefore, the process variance by (\ref{eqn:var_W}) is
\begin{align}
   v_{W}(\theta)&=\dfrac{1}{\theta^{2}}\bigg\{m_{2W}-m_{1W}^{2}+2\bigg[m_{1}(p,q) \bigg(\frac{p^{2}}{1-p}-q\bigg)-q\,\log q+\frac{p^{2}}{1-p}\,\text{log}(1-p)\bigg]\nonumber\\
   &\quad+\frac{p^{3}}{1-p}+q(1-q)+\frac{2p^{2}q}{1-p}\bigg \}\nonumber\\
   &:=\dfrac{1}{\theta^{2}}m_{3W}.
\end{align}

\subsection*{A.2 Pareto-Gamma}

The integral of the second moment quantile function of Pareto is
\begin{align*}
\int_{p}^{1-q}\big[(1-w)^{-1/t}-1\big]^{2}\,dw  
&=
-\int_{1-p}^{q}\big(u^{-1/t}-1\big)^{2}\,du 
=
-\int_{1-p}^{q}u^{-2/t}\,du+2\int_{1-p}^{q}u^{-1/t}\,du-\int_{1-p}^{q}1\,du \\
& =
-\left[\frac{t}{t-2} u^{\frac{t-2}{t}}\right]\bigg|_{1-p}^{q}+\left[\frac{2t}{t-1} u^{\frac{t-1}{t}}\right]\bigg|_{1-p}^{q}-u\bigg|_{1-p}^{q}\\&=\frac{t}{t-2}(1-p)^{\frac{t-2}{t}}-\frac{t}{t-2} q^{\frac{t-2}{t}}-\frac{2t}{t-1}(1-p)^{\frac{t-1}{t}}+\frac{2t}{t-1} q^{\frac{t-1}{t}}+(1-p)-q.
\end{align*}
\noindent
In equation (\ref{eqn:var_W}), the derivative of Pareto quantile functions is
\begin{align*}
& H^{'}(w)=\dfrac{1}{F^{'}\Big(\theta \big[(1-w)^{-1/t}-1\big]\Big)}=\dfrac{1}{t\theta^{t}/\Big[\theta+\theta[(1-w)^{-1/t}-1]\Big]^{t+1}}=\dfrac{\theta(1-w)^{-\frac{t+1}{t}}}{t}.
\end{align*}
\noindent
{\sc{Trimmed Version:}} \\
The hypothetical mean is
\begin{align*}
  \mu_{T}(\theta)&=\theta\bigg\{\frac{1}{1-p-q}\int_{p}^{1-q}\big[(1-w)^{-1/t}-1\big]dw\bigg\}  \nonumber\\
  &=\theta\,\frac{1}{1-p-q}\bigg\{(1-p)\Big[\frac{t}{t-1}(1-p)^{-1/t}-1\Big]-q\Big(\frac{t}{t-1}q^{-1/t}-1\Big)\bigg\} \nonumber\\
  &=\theta  m_{1T}.
\end{align*}
And the process variance is
\begin{align*}
v_{T}(\theta)&=\theta^2 \frac{1}{(1-p-q)^2}\int_{p}^{1-q}\int_{p}^{1-q}(\text{min}\{u,v\}-uv)\,\dfrac{(1-u)^{-\frac{t+1}{t}}}{t}\,\dfrac{(1-v)^{-\frac{t+1}{t}}}{t}\,du\,dv\\
& := \theta^2\,m_{3T}
 \end{align*}

\noindent
{\sc{Winsorized Version:}}\\
The hypothetical mean is
\begin{align*}
  \mu_{W}(\theta)&=\theta\bigg\{p\big[(1-p)^{-1/t}-1\big]+\int_{p}^{1-q}\big[(1-w)^{-1/t}-1\big]dw+q\big[q^{-1/t}-1\big]\bigg\}  \nonumber\\
  &=\theta\bigg[p\big[(1-p)^{-1/t}-1\big]+(1-p)\Big[\frac{t}{t-1}(1-p)^{-1/t}-1\Big]-q\Big(\frac{t}{t-1}q^{-1/t}-1\Big)+q\big(q^{-1/t}-1\big)\bigg] \nonumber\\
  &:=\theta  m_{1W}.
\end{align*}
\noindent
And the winsorized variance is
\begin{align*}
   \Var(X_{w}|\theta)
&=\theta^{2}\bigg\{p\big[(1-p)^{-1/t}-1\big]^{2}+ \int_{p}^{1-q}\big[(1-w)^{-1/t}-1\big]^{2}\,dw+q\big[q^{-1/t}-1\big]^{2}\bigg\}-[\mu_{W}(\theta)]^{2} \nonumber\\
 &=\theta^{2}\bigg\{p\big[(1-p)^{-1/t}-1\big]^{2}+\frac{t}{t-2}(1-p)^{\frac{t-2}{t}}-\frac{2t}{t-1}(1-p)^{\frac{t-1}{t}}+(1-p)\nonumber\\
&\qquad +q\big[q^{-1/t}-1\big]^{2}-\frac{t}{t-2}q^{\frac{t-2}{t}} +\frac{2t}{t-1} q^{\frac{t-1}{t}}-q\bigg\}-[\theta\,m_{1}]^{2} \nonumber\\
  &:=\theta^{2}\big(m_{2W}-m_{1W}^{2}\big).
\end{align*}

\noindent
Again, by (\ref{eqn:var_W}), the process variance becomes
\begin{align*}
v_{W}(\theta)&=\theta^{2}\bigg\{m_{2}(p,q)-m_{1}^{2}(p,q)+\frac{2}{t}\bigg[m_{1}(p,q) \Big(p^{2}(1-p)^{-\frac{t+1}{t}}-q^{\frac{t-1}{t}}\Big)+q^{\frac{t-1}{t}}\Big(q^{-\frac{1}{t}}-1\Big)-p^{2}(1-p)^{-\frac{t+1}{t}} \nonumber\\
& \quad \Big((1-p)^{-\frac{1}{t}}-1\Big)\bigg]+\frac{1}{t^{2}}\bigg[p^{3}(1-p)(1-p)^{-\frac{2(t+1)}{t}}+q^{3}(1-q)q^{-\frac{2(t+1)}{t}}+2p^{2}q^{2}(1-p)^{-\frac{t+1}{t}}q^{-\frac{t+1}{t}}\bigg]\bigg \}\nonumber\\
&:=\theta^{2}m_{3W}.
\end{align*}

\subsection*{A.3 Lognormal-Normal}

\noindent
The integrals of the Lognormal moment quantile with limited boundaries are \begin{align*}
\int_{0}^{a}F^{-1}(w)dw&=\int_{0}^{\pi_{a}}xf(x)dx=\int_{0}^{\pi_{a}}x\frac{\phi(z)}{\sigma^{'} x}dx=\int_{-\infty}^{\frac{\ln\pi_{a}-\theta}{\sigma^{'}}}\frac{\phi(z)}{\sigma ^{'}}d(e^{\theta+z\sigma}) \\
& =
e^{\theta+\frac{1}{2}\sigma^{'2}}\int_{-\infty}^{\frac{\ln\pi_{a}-\theta}{\sigma^{'}}}\frac{1}{\sqrt{2\pi}}e^{-(z-\sigma^{'})^2/2}dz
=
e^{\theta+\frac{1}{2}\sigma^{'2}}\int_{-\infty}^{\frac{\ln\pi_{a}-\theta}{\sigma^{'}}-\sigma^{'}}\frac{1}{\sqrt{2\pi}}e^{-t^2/2}dt \\ 
& =
e^{\theta+\frac{1}{2}\sigma^{'2}}\Phi\left(\Phi^{-1}(a)-\sigma^{'}\right) \\
\int_{0}^{a}\left[F^{-1}(w)\right]^2 dw 
& =
\int_{0}^{\pi_{a}}x^2 f(x)dx
=
\int_{-\infty}^{\frac{\ln\pi_{a}-\theta}{\sigma^{'}}}e^{\theta+z\sigma^{'}}\,\frac{\phi(z)}{\sigma^{'} }d(e^{\theta+z\sigma^{'}}) \\
& =
e^{2\theta}\int_{-\infty}^{\frac{\ln\pi_{a}-\theta}{\sigma^{'}}}e^{2z\sigma^{'}}\frac{1}{\sqrt{2\pi}}e^{-z^2/2}dz
=
e^{2\theta+2\sigma^{'2}}\int_{-\infty}^{\frac{\ln\pi_{a}-\theta}{\sigma^{'}}}\frac{1}{\sqrt{2\pi}}e^{-(z-2\sigma^{'2})}dz \\ 
& =
\int_{-\infty}^{\frac{\ln\pi_{a}-\theta}{\sigma^{'}}-2\sigma^{'}}\frac{1}{\sqrt{2\pi}}e^{-t^2/2}dt
=
e^{2\theta+2\sigma^{'2}}\,\Phi\left(\Phi^{-1}(a)-2\sigma^{'}\right).
\end{align*}
In equation (\ref{eqn:var_W}), the derivative of Pareto quantile functions is
\begin{align*}
  H^{'}(w)&=\dfrac{1}{F^{'}\big(e^{\theta+\sigma \Phi^{-1}(w)}\big)}=\dfrac{\sigma e^{\theta+\sigma\Phi^{-1}(w)}}{\frac{1}{\sqrt{2\pi}} e^{-(\frac{\ln e^{\theta+\sigma\Phi^{-1}(w)}-\theta}{\sigma})^2/2}}=\sqrt{2\pi}\sigma \, e^{[\Phi^{-1}(w)]^2/2+\theta+\sigma\Phi^{-1}(w)}.\end{align*}
{\sc Trimmed Version:}\\
The hypothetical mean is
\begin{align*}
  \mu_{T}(\theta)&=\frac{1}{1-p-q}\int_{p}^{1-q}e^{\theta+\sigma^{'}\Phi^{-1}(w)}dw  \nonumber\\
&=\frac{e^{\theta+\frac{1}{2}\sigma^{'2}}}{1-p-q}\left[\Phi\left(\Phi^{-1}(1-q)-\sigma^{'}\right)-\Phi\left(\Phi^{-1}(p)-\sigma^{'}\right)\right]\\
  &=e^{\theta}  m_{1T}.
\end{align*}
And the process variance is
\begin{align*}
v_{T}(\theta)
&= \frac{1}{(1-p-q)^2}\int_{p}^{1-q}\int_{p}^{1-q}(\text{min}\{u,v\}-uv)2\pi\sigma^2\,e^{[\Phi^{-1}(u)]^2/2+\theta+\sigma^{'}\Phi^{-1}(u)}e^{[\Phi^{-1}(v)]^2/2+\theta+\sigma^{'}\Phi^{-1}(v)}\,du\,dv\\
&:= 
e^{2\theta} m_{3T}.
 \end{align*}
\noindent
{\sc Winsorized Version:}\\
The hypothetical mean is
\begin{align*}
\mu_{W}(\theta)&=e^{\theta}\left\{p e^{\sigma^{'} \Phi^{-1}(p)}+\int_{p}^{1-q}e^{\sigma^{'}\Phi^{-1}(w)}dw+q e^{\sigma^{'}\Phi^{-1}(1-q)}\right\}  \nonumber\\
&=e^{\theta}\bigg\{p e^{\sigma^{'}\Phi^{-1}(p)}+e^{\frac{1}{2}\sigma^{'2}}\left[\Phi\left(\Phi^{-1}(1-q)-\sigma^{'}\right)-\Phi\left(\Phi^{-1}(p)-\sigma^{'}\right)\right]+q e^{\Phi^{-1}(1-q)}\bigg\}  \nonumber\\
&:=e^{\theta}  m_{1W}.
\end{align*}
And the winsorized variance is
\begin{eqnarray*}
\Var(X_{w}|\theta)
& = &
p \big[F^{-1}(p)\big]^{2}+ \int_{p}^{1-q}\big[F^{-1}(w)\big]^{2}\,dw+q\big[F^{-1}(1-q)\big]^{2}-[\mu_{W}(\theta)]^{2}
\nonumber \\
& = &
e^{2\theta}\left\{p e^{2\sigma^{'}\Phi^{-1}(p)}+ e^{2\sigma^{'2}}\left[\Phi\left(\Phi^{-1}(1-q)-2\sigma^{'}\right)-\Phi\left(\Phi^{-1}(p)-2\sigma^{'}\right)\right]
+q e^{2\sigma^{'}\Phi^{-1}(1-q)} \right\} \\
& & 
-e^{2\theta} m_{1W}^{2}\\
& := & e^{2\theta}\Big(m_{2W}-m_{1W}^{2}\Big).
\end{eqnarray*}
\noindent
Finally, the process variance is
\begin{align*}
v_{W}(\theta)
& =
e^{2\theta} 
\Bigg\{
m_{2W}-m_{1W}^{2}-\left(p^2 \Delta_{p}-q^2 \Delta_{1-q}\right)^2+p^3 \Delta_{p}^2 + q^3 \Delta_{1-q}^2  \\
& \hspace{12pt}
+2\bigg[ m_{1W} (p^2 \Delta_{p}-q^2 \Delta_{1-q} ) 
+q^2 \Delta_{1-q}\, e^{\sigma^{'}\Phi^{-1}(1-q)}
-p^2 \Delta_{p}\, e^{\sigma^{'}\Phi^{-1}(p)}\bigg] \Bigg\}
\nonumber\\
&:=e^{2\theta} m_{3W},
\end{align*}
where $\Delta_{p}=H^{'}(p)/e^{\theta}$ and $\Delta_{1-q}=H^{'}(1-q)/e^{\theta}$.

\subsection*{A.4 Non-parametric Model Property}

The empirical sample estimate of (\ref{eqn:var_T}) is
 \begin{align*}
\widehat{v}_{T}
& =
\frac{n^2}{(n-\floor*{np}-\floor*{nq})^2}
\sum_{j=\floor*{np}+1}^{n-\floor*{nq}} \
\sum_{i=\floor*{np}+1}^{n-\floor*{nq}}
\left(
\min\left\{\frac{j}{n},\frac{i}{n}\right\}-\frac{j}{n}\frac{i}{n}\right) \,
\left( 
x_{(j+1)}-x_{(j)}
\right) 
\left(
x_{(i+1)}-x_{(i)}
\right).
 \end{align*}
\noindent
The sample estimate of the components of (\ref{eqn:var_W}) are
\begin{align*}
& \widehat{\mu}_{W}
=
\frac{1}{n}
\left( 
\floor*{np} x_{\floor*{np}+1}+ \sum_{i=\floor*{np}+1}^{n-\floor*{nq}}x_{i}+\floor*{nq} x_{n-\floor*{nq}}
\right),
\\
&\widehat{\Var}(X_{W}|\theta)
= 
\frac{1}{n}
\left( 
\floor*{np} x_{\floor*{np}+1}^{2}+ \sum_{i=\floor*{np}+1}^{n-\floor*{nq}}x_{i}^{2}+\floor*{nq} x_{n-\floor*{nq}}^{2}
\right)
-\widehat{\mu}_{W}^{2},
\\
&\widehat{H}(p)= \begin{cases}
0.5(x_{np}+x_{np+1}); \quad \text{if $np$ is an integer}, \\
x_{\ceil*{np}}; \quad \text{if $np$ is not an integer},
\end{cases} \\
& 
\widehat{H}(1-q)
=
\begin{cases}
0.5(x_{n-nq}+x_{n-nq+1}); \quad \text{if $nq$ is an integer}, \\
x_{\ceil*{n-nq}}; \quad \text{if $nq$ is not an integer},
\end{cases} \\ 
&\widehat{A} = \frac{(\floor*{np})}{n}^{2} \big(x_{\ceil*{np}+1} -  x_{\ceil*{np}}\big);
\qquad
\widehat{B} = \frac{(\floor*{nq})}{n}^{2} 
\left(
x_{\ceil*
{n-nq}} -  x_{\ceil*{n-nq}-1}
\right).
\end{align*}
\end{document}